\documentclass[letterpaper,12pt,final,openany]{aastex631}
\usepackage{amssymb,amsmath,amsfonts}
\usepackage[utf8x]{inputenc}

\usepackage{color}

\def\kms{km~s$^{-1}$}

\newcommand{\vsini}{\ensuremath{v_{{\mathrm e}}\sin i}}
\newcommand{\teff}{\ensuremath{T_\mathrm{eff}}}

\newcommand{\logg}{\ensuremath{\mathrm{log}\ g}}

\received{}
\revised{}
\accepted{}

\shorttitle{Six newly discovered CP stars}
\shortauthors{Monier et al.}

\begin{document}


\title{The surface composition of six newly discovered chemically peculiar stars.\\ Comparison to the HgMn stars $\mu$ Lep and $\beta$ Scl and the superficially normal B star $\nu$ Cap}

\author{Richard Monier}
\affiliation{LESIA, UMR 8109, Observatoire de Paris et Universit\'e Pierre et Marie Curie Sorbonne Universit\'es, place J. Janssen, Meudon, France.}

\author{E. Niemczura}
\affiliation{University of Wroclaw, Astronomical Institute, Kopernika 11,PL-51-622, Wroclaw, Poland}

\author{D.W. Kurtz}
\affiliation{Department of Physics, North-West University, Dr Albert Luthuli Drive, Mahikeng 2735, South Africa}
\affiliation{Jeremiah Horrocks Institute, University of Central Lancashire, Preston PR1 2HE, UK}

\author{S.Rappaport}
\affiliation{Department of Physics, Kavli Institute for Astrophysics and Space Research, M.I.T., Cambridge, MA 02139, USA}

\author{D.M. Bowman}
\affiliation{Institute of Atronomy, KU Leuwen, Celestijnenlaan 200D, B-3001 Leuven, Belgium}

\author{Simon J. Murphy}
\affiliation{University of Southern Queensland, Centre for Astrophysics, West Street, Toowoomba, QLD 4350 Australia}

\author{Yveline Lebreton}
\affiliation{LESIA, UMR 8109, Observatoire de Paris et Universit\'e Pierre et Marie Curie Sorbonne Universit\'es, place J. Janssen, Meudon, France.}
\affiliation{Univ Rennes, CNRS, IPR (Institut de Physique de Rennes) - UMR 6251, F-35000 Rennes, France.}

\author[0000-0001-7797-3749]{Remko Stuik}
\affiliation{Leiden Observatory, Leiden University, Niels Bohrweg 2, 2300 RA, Leiden, The Netherlands.}

\author{Morgan Deal}
\affiliation{Instituto de Astrof\'isica e Ci\^encias do Espaço, Universidade do Porto, CAUP, Rua das Estrelas, PT4150-762 Porto, Portugal}

\author{Thibault Merle}
\affiliation{Institut d'astronomie et d'astrophysique, Universit\'{e} Libre de Bruxelles, CP. 226, Boulevard du Triomphe, 1050 Brussels, Belgium}

\author{T. K{\i}l{\i}\c{c}o\u{g}lu}
\affiliation{Department of Astronomy and Space Sciences, Faculty of Science, Ankara University, 06100, Turkey.}

\author[0000-0002-8675-4000]{M. Gebran}
\affiliation{Department of Chemistry and Physics, Saint Mary's College, Notre Dame, IN 46556, USA.}

\author{Ewen Le Ster}
\affiliation{Univ Rennes, CNRS, IPR (Institut de Physique de Rennes) - UMR 6251, F-35000 Rennes, France.}

\correspondingauthor{Richard Monier}
\email{Richard.Monier@obspm.fr}




\begin{abstract}
We report on a detailed abundance study of six bright, mostly southern, slowly rotating late B stars: HD~1279 (B8III), HD~99803 (B9V), HD~123445 (B9V), HD~147550 (B9V), HD~171961 (B8III) and  HD~202671 (B5II/III), hitherto reported as normal stars. We compare them to the two classical HgMn stars $\mu$ Lep and $\beta$ Scl and to the superficially normal star, $\nu$ Cap. In the spectra of the six stars, the \ion{Hg}{2} line at 3984 \AA\ line is clearly seen and numerous lines of P, Ti, Mn, Fe, Ga, Sr, Y, and Zr appear to be strong absorbers. 
A comparison of newly acquired and archival  spectra of these objects with a grid of synthetic spectra for selected unblended lines reveals large overabundances of P, Ti, Cr, Mn, Sr, Y, Zr, Ba, Pt and Hg and underabundances of He, Mg, Sc and Ni.
The effective temperatures, surface gravities, low projected rotational velocities and the peculiar abundance patterns of the six investigated stars show that they are new chemically peculiar stars, mostly new HgMn stars, and are reclassified as such. 
  The evolutionary status of these stars has been inferred and their ages and masses estimated. The two most massive objects, HD~1279 and  HD~202671, might have evolved away from the main-sequence recently, the other stars are main-sequence objects. HD~99803A is a sharp lined HgMn star with grazing eclipses; from TESS and MASCARA photometry we determine an orbital period of $P_{\rm orb} = 26.12022 \pm 0.00004$\,d.  
\end{abstract}

\keywords{stars: chemically peculiar --- stars: individual (HD~1279, HD~99803, HD~123445, HD~147550, HD~171961,  HD~202671, $\mu$ Lep, $\beta$ Scl, $\nu$ Cap)}

\section{Introduction}

The bright southern stars \textbf{HD~99803 (HR~4423, B9V, $V=5.14$), HD~123445 (HR~5294, B9V, $V=6.17$), HD~147550 (HR~6096, B9V, $V=6.24$), HD~171961 (HR~6990, B8III, $V=5.77$) and  HD~202671 (HR~8137, B5II/III, $V=5.38$)} have received little attention for their brightnesses: only 38, 57, 93, 35 and 79 references respectively can be found in SIMBAD for these stars.  

We have recently undertaken a spectroscopic survey of all apparently slowly rotating late B stars (B8 -- B9V) brighter than magnitude 7.0 and observable in the southern hemisphere which have $v_{\rm e} \sin i \le 60$\,km~s$^{-1}$ and $\delta$ $\le$ $-10^{\circ}$. We have added to these objects the northern star HD~1279 (HR~62, B8III, 53 references in SIMBAD) for which we obtained discretionary time at Observatoire de Haute Provence in January 2019. 

This project addresses some fundamental questions of the physics of late-B stars: 
i) Is our census of Chemically Peculiar (CP) stars complete up to the magnitude limits we adopted? If not, what are the  physical properties of the newly found CP stars? ii) Can we find new instances of rapid rotators seen pole-on (other than Vega) and study their physical properties (gradient of temperature across the disk, limb and gravity darkening)?
 
\cite{Royer} carried out a similar exercise on a sample of 47 apparently slowly rotating A0 -- A1V stars in the northern hemisphere which satisfy $v_{\rm e} \sin i \le 60$\,km~s$^{-1}$ and $\delta$ $\ge$ $-10^{\circ}$. Abundance analyses of selected lines allowed them to sort these stars into 3 groups: 13 chemically peculiar stars (among which 4 were new CPs), 17 superficially normal stars and 17 spectroscopic binaries (\citealt{Royer}). \cite{Monier} reported on a first analysis of the abundances of Fe, Mn and Hg and showed that four members of that sample: HD~30085, HD~18104, HD~32867 and HD~53588 are four newly discovered HgMn stars. \cite{2018ApJ...854...50M} have shown that HR 8844 is a new CP star, possibly an intermediate object between the Am and HgMn stars. \cite{2019AJ....158..157M} provide a detailed abundance analyses of HD~30085 and HD~30963 which highlight the HgMn status of these 2 stars.

In this paper we report on abundance analyses from high resolution, high signal-to-noise, optical range \'echelle spectra of HD~1279 and five stars of the southern sample, HD~99803, HD~123445, HD~147550, HD~171961 and  HD~202671. We also derive elemental abundances \textbf{of $\mu$ Lep (HR~1702, HD~33904) and $\beta$ Scl, (HR~8937, HD~221507)} two classical HgMn stars, which we use as comparison stars. We also use \textbf{$\nu$ Cap (HR~7773, HD~193432)}, a superficially normal B9~V star, as a comparison.

Using model atmospheres and line synthesis, we  derive for the first time the abundances of up to forty chemical elements in these six stars and find that they depart strongly from solar. The overabundances of Mn, Sr, Y, Zr, Pt and Hg and the underabundances of 
He, Mg, Sc and Ni establish that these stars are newly discovered CP stars, mostly new HgMn or PGa stars, whose atmospheres respectively harbour overabundances of manganese and mercury and overabundances of phosphorus and gallium. 

The paper is divided into seven sections. The first section recapitulates results from previous spectroscopic work on \textbf{HD~1279 (HR~62), HD~99803 (HR~4423), HD~123445 (HR~5294), HD~147550 (HR~6096), HD~171961 (HR~6990) and  HD~202671 (HR~8137)}, the second section describes new or archival spectroscopic observations of the eight stars. 
In the third section, we present the derivation of fundamental parameters, the modelling of the evolution of these new CP stars, including radiative diffusion and rotational mixing, and we provide estimates of their masses and ages.
In the fourth section, we present frequency analyses of the recent lightcurves of these stars obtained with
the Transiting Exoplanet Survey Satellite, \cite{2015JATIS...1a4003R}
and we derive a rotational period when feasible.
In section 5, we discuss the model atmosphere analyses and  the computations of synthetic spectra and, in section 6, the determination of elemental abundances for each star. In the conclusion, we discuss the chemical peculiarity of the six newly discovered chemically peculiar stars in the light of what is known of other HgMn stars.

\section{Previous spectroscopic work on HD~1279, HD~99803, HD~123445, HD~147550, HD~171961,  HD~202671, $\mu$ Lep and $\beta$ Scl}

HD~1279 (HR~62) was ascribed a spectral type B7II by \cite{1994AJ....107.1556G}. This star does not have a Renson nor a CCDM  number: it was not previously known to be a chemically peculiar star or a spectroscopic binary. HD~99803 (HR~4423) is a visual binary with 13\,arcsec separation. \citet{1984ApJS...55..657C} classified HD~99803A as B9IV and HD~99803B as Am: kA3hA5VmA7 (meaning a marginal Am star); \citet{2020RNAAS...4..234M} found HD~99803A to be a new HgMn star. HD~99803B, is fainter ($V=7.77$) and cooler \textbf{than} HD~99803A;  the lines of the two components are clearly separated in the HARPS spectrum we have used). HD~123445 (HR~5294) is CCDM J14089-4328A, the other member of the system is much fainter and cooler ($V=12.52$). We see no signs of this companion in the spectra we analysed. HD~123445A (HR~5294)  does not have any Renson member. HD~171961 (HR~6990) has neither a CCDM nor a Renson number. 
HD~147550 (HR~6096) is Renson 41660, it is classified as a B9 He-weak star.  HD~202671 (HR~8137) is Renson 56480 and is classified as a B7 He-weak Mn star. \cite{2014PASJ...66...23T} derived the abundance of sodium and iron for this star, and  \cite{1997A&A...320..257L} derived the effective temperature. 
No extensive abundance study (i.e., for several elements) has been published for any of these six stars.

In contrast, the abundances of $\mu$ Lep (HR~1702) and $\beta$ Scl (HR~8937) have been well studied. General abundance analyses of $\mu$ Lep and $\beta$ Scl can be found in \cite{1997A&AS..125..219A}
and \cite{2010MNRAS.405.1384T}, respectively. Other works focus on the determination of the abundance of one element in several HgMn stars, they are collected in Table \ref{tab:abund}.

\begin{table}[!h]
\caption{Previous abundance studies for $\mu$ Lep (HR~1702) and $\beta$ Scl}
\label{tab:abund}
\centering
\begin{tabular}{||c|c|c|}
\hline  \hline 
Chemical element(s)   &  Reference & star  \\ \hline      
Sr, Y       &  \cite{2008MNRAS.385.1523D}& $\mu$ Lep and $\beta$ Scl  \\
Xe           &  \cite{2008MNRAS.385.1523D} & $\mu$ Lep and $\beta$ Scl  \\
Ca           &  \cite{2007MNRAS.377.1579C} & $\mu$ Lep      \\
             &  \cite{Castelli2004} & $\beta$ Scl   \\
Ga           &  \cite{2005AJ....130.2312N} & $\mu$ Lep      \\
             & \cite{Dwo98} & $\mu$ Lep and $\beta$ Scl \\ 
	     &  \cite{Smith96}    & $\mu$ Lep and $\beta$ Scl \\
	     & \cite{Lanz1993}  & $\beta$ Scl   \\
Hg           & \cite{Do03}& $\mu$ Lep and $\beta$ Scl \\
             & \cite{1999ApJ...521..414W} & $\mu$ Lep  \\ 
	     &\cite{Smith97}  & $\beta$ Scl   \\
Cr           &  \cite{2003ASPC..305..230S} & $\mu$ Lep   \\
Ne           & \cite{2000MNRAS.318.1264D} & $\mu$ Lep   \\
N            & \cite{1999ApJ...524..974R}  & $\mu$ Lep     \\
Mn           & \cite{1998CoSka..27..324J} & $\mu$ Lep and $\beta$ Scl \\
O            & \cite{1999PASJ...51..961T})  & $\mu$ Lep     \\
Cu, Zn       & \cite{Smith94}       & $\mu$ Lep and $\beta$ Scl \\
Fe-peak      & \cite{smith1993}& $\mu$ Lep and $\beta$ Scl \\
             & \cite{1993ApJ...419..276A}       & $\mu$ Lep  \\
Mg, Al, Si   &  \cite{smith1993}       & $\mu$ Lep and $\beta$ Scl \\
C, N, O      &\cite{1990ApJS...73...67R}      & $\mu$ Lep      \\
C            & \cite{2002NewA....7..495C}     & $\beta$ Scl    \\             
\hline	\hline
\end{tabular}
\end{table}

\section{Observations}


For the comparison stars $\mu$ Lep (HR~1702) and $\beta$ Scl (HR~8937), we have used archival FEROS spectra (Fiber-fed Extended Range Optical Spectrograph , R = 48000 \footnote{\url{https://www.eso.org/sci/facilities/lasilla/instruments/feros.html}}) with good signal-to-noise ratios.
The spectra that we use were retrieved from the ESO Spectral archive \footnote{\url{http://archive.eso.org/wdb/wdb/adp/phase3_spectral/form}}.

For the northern star HD~1279 (HR~62), one high resolution spectrum was obtained in service observing
at Observatoire de Haute Provence with SOPHIE \footnote{http://www.obs-hp.fr/guide/sophie/sophie-eng.shtml}, the \'echelle spectrograph in its high resolution mode ($R=75000$) yielding a full spectral coverage from 3820 \AA\ to 6930 \AA\ in 39 orders.
A detailed technical description of SOPHIE is given in \cite{Perruchot}. SOPHIE is a cross-dispersed, environmentally stabilized \'echelle spectrograph dedicated to high precision radial velocity measurements. The spectra are extracted online from the detector images using a specific pipeline adapted from that of HARPS
(High Accuracy Radial Velocity Planet Searcher \footnote{https://www.eso.org/sci/facilities/lasilla/instruments/harps.html}).
  
HD\,171961 (HR~6990) was observed five times using the High Resolution Spectrograph (HRS\footnote{https://astronomers.salt.ac.za/instruments/hrs/}, \citealt{SALT14}) behind the 11-m Southern African Large Telescope (SALT \footnote{https://astronomers.salt.ac.za/}). We have coadded the individual spectra.
The SALT HRS is a dual-beam (370-550 nm \& 550-890 nm) fibre-fed, white-pupil, \'echelle spectrograph, employing VPH gratings as cross dispersers.  The cameras are all-refractive.  SALT HRS is an efficient single-object spectrograph using pairs of large (350 $\mu$m and 500 $\mu$m; 1.6 and 2.2 arcsec) diameter optical fibers, one for the star and one for background sky.  Three of the pairs feed image slicers before injection into the spectrograph, which delivers in its highest resolution mode a resolving power of $R \simeq 65,000$ (sliced 350 $\mu$m fibres).  A single 2k $\times$ 4k CCD captures all of the blue orders, while a 4k $\times$ 4k detector, using a fringe-suppressing deep-depletion CCD, is used for the red camera. 

For the four remaining stars, HD\,99803 (HR~4423), HD\,123445 (HR~294), HD\,147550 (HR~6096) and HD\,202671 (HR~8137), we have used FEROS or HARPS archival spectra. 

The observing dates, exposure times and signal-to-noise ratios at 5000 \AA\ of the spectra are collected in Table \ref{tab:obs}.  When several spectra were available for one star, we have used the spectrum with the highest signal-to-noise ratio. In all cases, we have used the one dimensional spectrum provided by the project.


\begin{table}[!h]
\caption{Log of observed and archival spectra}
\label{tab:obs}
\centering
\begin{tabular}{||c|c|c|c|c|c|}
\hline \hline
Star ID      & Observation& Instrument & R      & Exposition      & S/N \\ 
             & Date       &            &        & Time [s]        &     \\ \hline
  $\mu$\,Lep & 08/12/2011 & FEROS      & 48000  & 60              & 517 \\
             & 21/12/2019 & HRS        & 65000  & 60(V)+50(NIR)   & 450 \\
 $\beta$\,Scl & 21/10/2005 & FEROS      & 48000  & 60              & 387 \\
             & 21/12/2019 & HRS        & 65000  & 120(V)+80(NIR)  & 400\\
 HD\,1279 & 14/01/2019 & SOPHIE     & 75000  & 1200            & 209 \\
  HD\,99803 & 29/04/2019 & HRS        & 65000  & 144(V)+118(NIR) & 200 \\                  
            & 07/04/2009 & HARPS      & 115000 & 200             & 305 \\
 HD\,123445  & 02/02/2010 & FEROS      & 48000 & 450 & 277   \\                 
 HD\,147550 & 24/03/2006 & UVES     & 40970 & 25 & 278 \\ 
            &  24/03/2006 & UVES    & 40970 & 25 &  278   \\    
 HD\,171961 & 26/07/2019 & HRS        & 65000  & 206(V)+155(NIR) & 220\\
	    & 27/07/2019 & HRS        & 65000  & 206(V)+155(NIR) & 320\\
	    & 28/07/2019 & HRS        & 65000  & 206(V)+155(NIR) & 190 \\
	    & 18/08/2019 & HRS        & 65000  & 206(V)+155(NIR) & 230 \\
	    & 19/08/2019 & HRS        & 65000  & 206(V)+155(NIR) & 240\\
HD\,202671 & 21/08/2013 & FEROS      & 48000 & 600 & 457 \\          
\hline \hline
\end{tabular}
\end{table}

\section{Evolutionary status and rotational periods determination}

\subsection{Fundamental parameters}

The fundamental data for HD~1279 (HR~62), HD~99803A (HR~4423), HD~123445 (HR~5294), HD~147550 (HR~6096), HD~171961 (HR~6990),  HD~202671 (HR~8137), $\beta$~Scl (HR~8937) and $\mu$~Lep (HR~1702) are collected in Table \ref{tab:fundamentals}. 
The spectral type retrieved from SIMBAD appears in row 2 and the apparent magnitudes in row 3, the Str\"omgren indexes $b -y$, $m_{1}$ and $c_{1}$ in rows 4, 5, 6.  The photometric data were taken from \cite{1998A&AS..129..431H}.


For all  stars, the effective temperature (\teff) and surface gravity (\logg) were determined  using the UVBYBETA code developed by \cite{Napiwotzki}. They are collected in rows 7 and 8. This code is based on the \cite{1985MNRAS.217..305M} grid, which calibrates the $uvby\beta$ photometry in terms of \teff \ and \logg. The estimated errors on \teff \ and \logg, are $\pm$125 K and $\pm$0.20 dex, respectively (see Sec. 4.2 in \citealt{Napiwotzki}). 

The radial velocities were derived by cross-correlating \ion{Fe}{2} lines with very accurate wavelengths from NIST in the range $4500-4600$ \AA\  with a synthetic spectrum computed for the derived effective temperature and surface gravity of the star and solar abundances. A parabolic fit to the upper part of the resulting cross-correlation function yields the Doppler shift, which is then used to shift spectra to rest wavelengths. The projected rotational velocities were derived from the position of the first zero of the Fourier transform of individual lines; they are taken from \cite{Royer}. The radial velocity, the microturbulent velocity and projected equatorial velocity are collected in rows 8, 9 and 10 in Table \ref{tab:fundamentals}. 
\\



\begin{table}
\caption{Adopted fundamental parameters for HD~1279, HD~99803A, HD~123445, HD~147550, HD~171961,  HD~202671, $\beta$~Scl and $\mu$~Lep. Uncertainties are $125$~K on $T_{\rm eff}$, $0.20$~dex on \logg~ and $0.10$ dex on the present surface metallicity $\mathrm{[Fe/H]_s}$. MS stands for main-sequence and SG for sub-giant.}
\label{tab:fundamentals}
\centering
\begin{tabular}{ll|c|c|c|c|c|c|c|c}
	\hline  \hline
& & HD~1279 & HD\,99803 & HD\,123445 & HD\,147550 & HD\,171961 & HD\,202671 & $\beta$~Scl & $\mu$~Lep  \\
\hline
Sp.T. & & B8III & B9V & B9V & B9V & B8III & B5II/III & B9.5III & B9IV\\
$V$ & & 5.852 & 5.136 & 6.170 & 6.241 & 5.772 & 5.383 &  4.370 & 3.290\\
$b-y$ & & -0.020 & -0.012 & -0.016& 0.059& 0.032& -0.048& -0.046& -0.043\\
$m_1$ & & 0.092 & 0.122 & 0.109 &0.109  & 0.074& 0.096& 0.125& 0.106 \\
$c_1$ & & 0.588 & 0.944 & 0.746 & 1.035 & 0.558 & 0.547& 0.677& 0.653 \\
$T_{\rm{eff}}$ & [K] & $13300$ & $10700$ & $12000$ & $10700$ & $13800$ & $13700$ & $12500$ & $12800$ \\
log\,$g$ & (cgs) & $3.16$ & $3.85$ & $4.19$ & $3.83$ & $3.85$ & $3.30$ & $4.13$ & $3.77$ \\
$v_{\rm e} \sin i$  & [km\,s$^{-1}$] & 28.0& 8.0& 66.0 & 4.8 & 55.0 & 25.0 & 26.0& 18.0 \\
$v_{\rm{micr.}}$ & [km\,s$^{-1}$] & 0.0 & 0.1 & 0.2 & 0.70 & 0.0 & 0.9 & 0.0 & 0.0 \\
$v_{\rm{rad}}$ & [km\,s$^{-1}$] & -9.20 & 3.0 & -2.0 & -24.0 & -33.0 & -12.50 & 0.40 & 27.10\\
$\mathrm{[Fe/H]_s}$ && $ 0.18$ & $-0.44$ & $-0.52$ & $ 0.05$ & $ 0.00$ & $ 0.40$ & $ 0.23$ & $ 0.23$ \\ 
$M_\mathrm{SPInS}$ & [M$_\odot$]  & $4.8 \pm 0.4$ &  $2.8 \pm 0.3$ & $3.0 \pm 0.2$ & $2.8 \pm 0.3$ & $4.0 \pm 0.4$ & $4.9 \pm 0.4$ & $3.2 \pm 0.2$ & $3.7 \pm 0.4$ \\
$A_\mathrm{SPInS}$ & (Myr) & $106 \pm 14$ & $260 \pm 63$ & $119 \pm 60$ & $261 \pm 62$ & $100 \pm 26$ & $100 \pm 12$ & $110 \pm 51$ & $138 \pm 28$ \\
Phase & & SG/MS & MS & MS & MS & MS & SG/MS & MS & MS  \\
\end{tabular}
\end{table}

\subsection{Evolutionary status and mass and age determinations}

\begin{figure}[h]
\vskip 0.5cm
   \centering
      \includegraphics[scale=0.80]{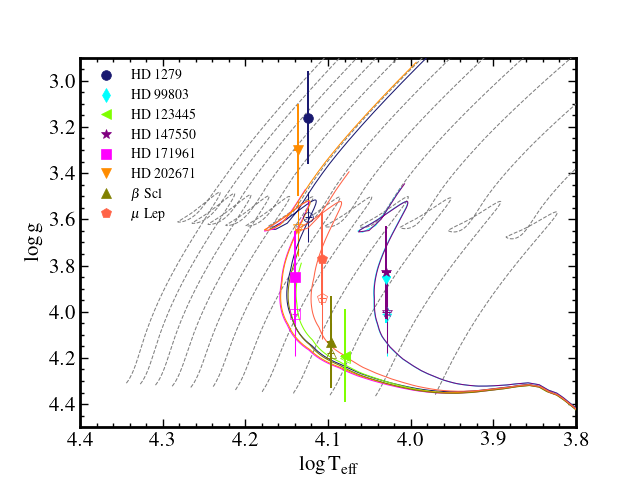}
      \caption{Positions of the six newly discovered chemically peculiar stars, and $\mu$ Lep and $\beta$ Scl in an effective temperature-surface gravity Kiel diagram. Full symbols show the positions of the stars  corresponding to the observed values of $\log g$ and $T_\mathrm{eff}$ (see Table \ref{tab:fundamentals}), while open symbols are for the best stellar model values inferred by SPInS. Colored lines are isochrones calculated for the age $A_\mathrm{SPInS}$ inferred by SPInS. Grey dashed lines are BaSTI-IAC evolutionary tracks of solar metallicity with masses increasing, from right to left, from $2.0$ to $7.5\, M_\odot$, by steps of $0.5\,M_\odot$.}
     \label{Fig:Kiel}\vskip 0.5cm
\end{figure}

To estimate the mass and age of the stars, we used the SPInS stellar model optimization tool \citep{Lebreton2020L}.  SPInS uses a Bayesian approach to find the probability distribution function of stellar parameters from a set of constraints. At the heart of the code is a Markov Chain Monte Carlo solver coupled with interpolation within a pre-computed stellar model grid. Here, we used the BaSTI-IAC grid of stellar models \citep{Hidalgo2018}. This grid is for a solar-scaled heavy element distribution with the solar mixture taken from \citet{Caffau2011} complemented by \citet{Lodders2010}, which corresponds to $(Z/X)_\odot=0.0209$. The grid considers convective core overshooting included as an instantaneous mixing between Schwarzschild's convective limit up to layers at a distance $\alpha_\mathrm{ov}=0.2 H_P$ from it, where $H_P$ is the pressure scale height at the Schwarzschild limit. Neither microscopic diffusion, nor rotation are taken into account. We considered as prior, \citet{Salpeter1955}'s initial mass function.

We estimate the age and mass of each star on the basis of the Kiel diagram observational constraints, 
that is effective temperature $T_\mathrm{eff}$ and surface gravity $\log g$.
We assume that the initial metallicity of the stars is close to solar, i.e. $\mathrm{[Fe/H]=0.0 \pm 0.10}$ dex. 
We provide the inferred mean ages $A_\mathrm{SPInS}$ and masses $M_\mathrm{SPInS}$ and their $1\sigma$ errors in Table \ref{tab:fundamentals}. The positions of the stars in the Kiel diagram are plotted in Fig. \ref{Fig:Kiel}, together with isochrones of age and metallicity inferred by SPInS.
The figure also shows BaSTI-IAC stellar evolutionary tracks of solar metallicity with masses increasing, from right to left, from $2.0$ to $7.5\, M_\odot$, by steps of $0.5\,M_\odot$. 

The six stars are found to be young with ages between $100$ and $261$ Myr and masses in the range $2.8-4.9\, M_\odot$. For each couple of stars (HD~99803A (HR~4423), HD~147550 (HR~6096)) and (HD~123445 (HR~5294), $\beta$~Scl (HR~8937)), stars have very close Kiel diagram parameters so that their ages and masses are very close. These stars lie on the main sequence and so do HD~171961 (HR~6990) and $\mu$~Lep (HR~1702). For stars HD~1279 (HR~62) and HD~202671 (HR~8137), the situation is less clear. Considering the observed parameters, the stars are expected to be on the subgiant branch but the SPInS inference places them close to the terminal age main sequence (TAMS) with surface gravities lower than the observed ones and error bars that do not overlap.
More generally, it is interesting to note that for all stars but two (HD~123445 (HR~5294) and $\beta$~Scl (HR~8937)), SPInS can find a solution corresponding to the observed effective temperature but for a $\log g$-value  higher than the observed one. It is difficult to identify the exact reason of the disagreement. On the theoretical side, these rather massive stars have very probably experienced rotation in the previous phases of their evolution, as well as microscopic diffusion and radiative accelerations on chemical elements. Rotation and gravitational settling bring fresh fuel to the H-burning core which impacts the age-dating and enlarges the main sequence. Concerning rotation, it has been shown by \citet{Ekstrom2012} that rotation combined with a moderate amount of overshooting ($\alpha_\mathrm{ov}=0.1 H_P$) have similar effects on the main sequence width than \textbf{compared with overshooting alone}  but with $\alpha_\mathrm{ov}=0.2 H_P$ which is the amount considered in the BaSTI grids. Concerning diffusion, we calculated the evolution on the MS of a star of mass $M=5.0\ M_\odot$ with the cestam code including microscopic diffusion and radiative accelerations \citep{morel08,marques13,deal18} and checked that it has no impact for these stars. A precise modeling of each star is very demanding and is therefore beyond the scope of this work, in particular because, on the observational side, the error on the values of $\log g$ is quite high which hampers a precise characterization of these stars.

\section{Analysis of photometric data}

\subsection{Analysis of the photometric TESS and MASCARA data of HD~99803 (HR~4423)}

\subsubsection{Frequency analysis of the TESS data- HD~99803 (HR~4423)}
\label{sec:fa1}

HD~99803 (TIC\,163204899; HR\,4423) was observed by {\em TESS} in Sector 10 in 2-min cadence. The data are available in both SAP (simple aperture photometry) and PDCSAP (presearch-data conditioning SAP); we used the PDCSAP data for our analysis, converting intensity to magnitudes. The data have a time span of 24.96\,d with a centre point in time of $t_0 = {\rm BJD}~2458583.20612$, and comprise 15992 data points, after clipping 51 outliers, which are probably the result of cosmic ray strikes (including solar radiation). 

There is also a 2.6-d gap in the data at the time of perigee. The {\em TESS} orbit is a lunar-synchronous eccentric orbit with an orbital period of half the sidereal orbital period of the moon. This keeps the satellite from being perturbed significantly at apogee, and allows for higher bandwidth data downloads at perigee. The gap in the data thus comes during the data transfers. 

\begin{figure}
\begin{center}
\includegraphics[width=1.0\linewidth,angle=0]{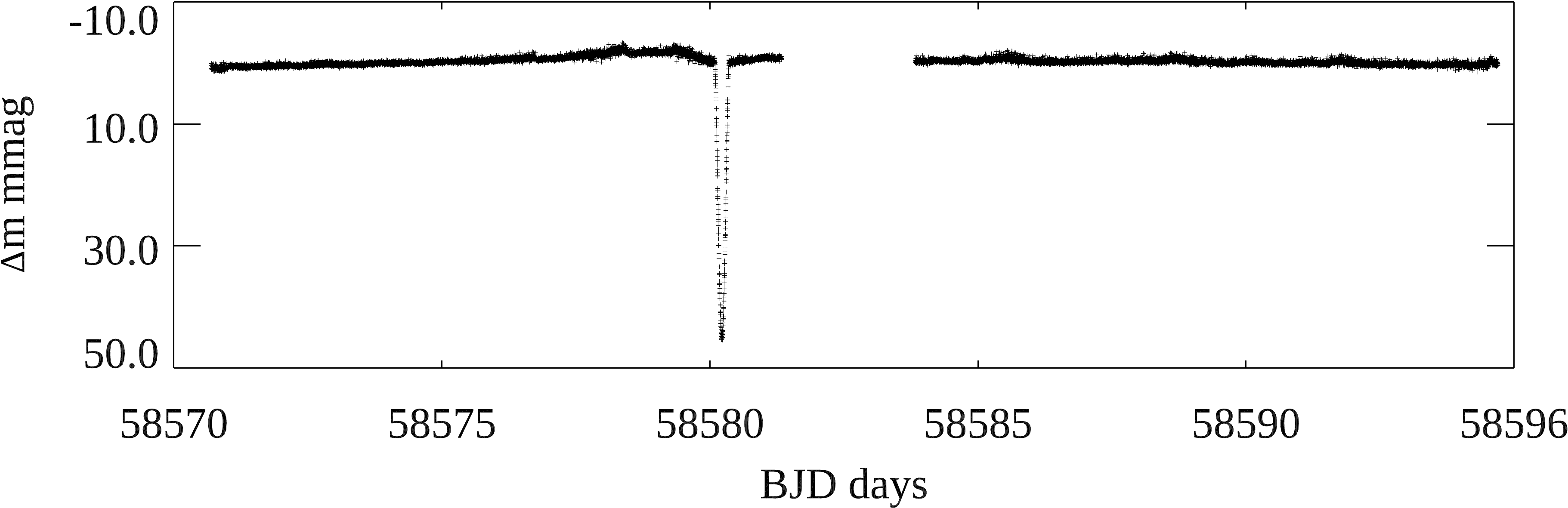}
\includegraphics[width=1.0\linewidth,angle=0]{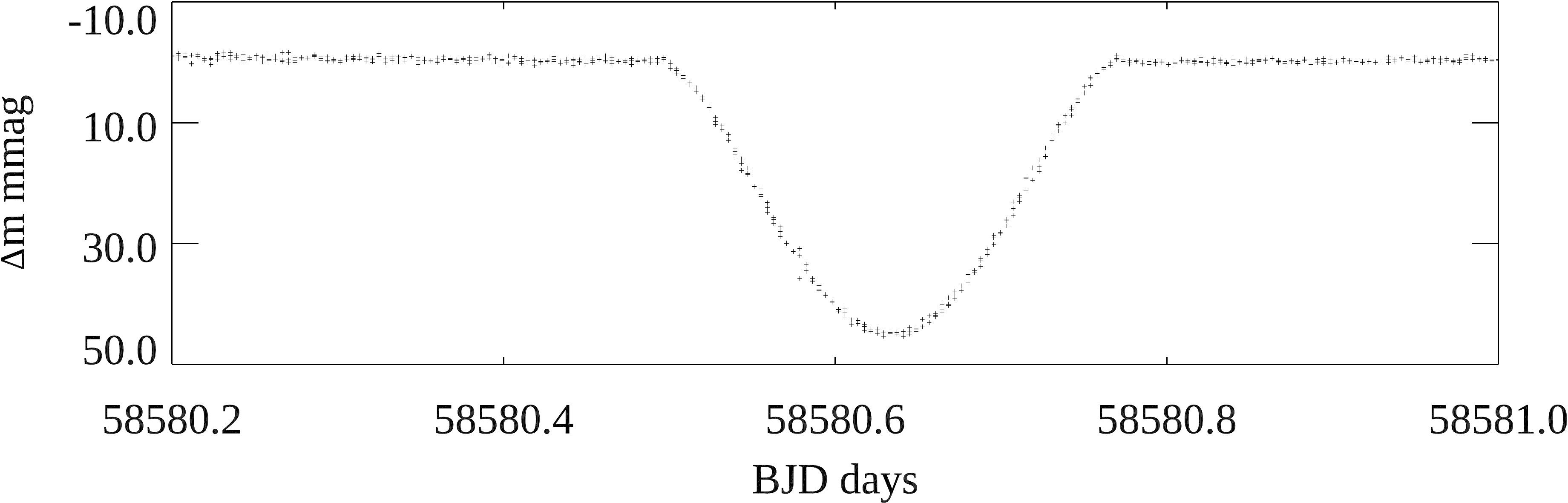}
\caption{The light curve of HD~99803 obtained in 120-s cadence in Sector 10. The ordinate scale is Barycentric Julian Date -- 240\,0000.0. The top panel is the full light curve; the bottom panel shows a higher time resolution view of the eclipse. 
}
\label{fig:lc} 
\end{center}
\end{figure}

Fig.\,\ref{fig:lc} shows the light curve of the Sector 10 data where an eclipse is obvious. This eclipse was also noted by \citet{2021MNRAS.506.5328K}. 
The bottom panel of the figure shows a higher time resolution of the eclipse. HD~99803 (HR~4423) was observed again by {\em TESS} in Sector 36, two years later, in 10-min cadence, and another eclipse was observed. The S36 data are from Full Frame Images and the star was close to an edge in this sector, making extraction of the data problematic. Nevertheless, with assistance from Tibor Mitnyan (private communication) who used Andras P\'al's FITSH package \citep{Pal12}, we were able to extract a time for the eclipse. We discuss the eclipses and the orbital period inferred from them in the next subsection.

\begin{figure}
\begin{center}
\includegraphics[width=1.0\linewidth,angle=0]{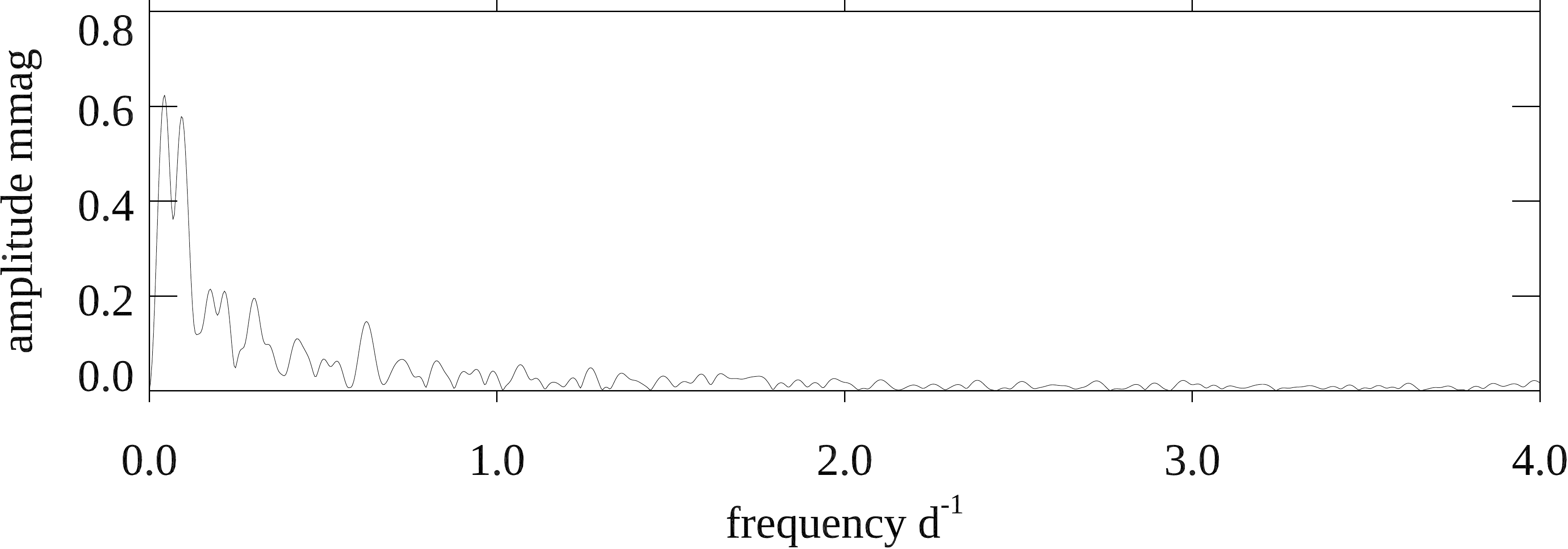}
\caption{An amplitude spectrum of the Sector 10 data for HD~99803 (HR~4423) after the eclipse was excised from the time string. There is no evidence of any pulsation out to the Nyquist frequency near 360\,d$^{-1}$. This plot shows on the low frequency range where there is a possibility of g~mode frequencies in a B9V star. No peaks that cannot be ascribed to instrumental variations are seen.}
\label{fig:ft} 
\end{center}
\end{figure}

The S10 data were analysed using a fast Discrete Fourier Transform \citep{1985MNRAS.213..773K} to produce an amplitude spectrum (Fig.\,\ref{fig:ft}).  There is no evidence of any p~mode pulsations out to the Nyquist frequency near 360\,d$^{-1}$ (not shown). In the low frequency range where g~modes occur there is no peak that cannot be ascribed to instrumental variations. Hence there is also no evidence of g~modes. Comparison with the instability strips of \citet{2008JPhCS.118a2079Z} show that HD~99803A (HR~4423) is too cool to be either a $\beta$\,Cep star or an SPB star. There is a small peak at 0.6\,d$^{-1}$ that may appear to be of interest, but when dividing the data into two subsets, this peak is most evident in the second half of the data set, \textbf{so} it is not convincing as a pulsation frequency.

HD~99803A and B fall within one pixel in TESS observations. The problem of HD~99803A lying near to a chip edge in the S36 FFI data may indicate that HD~99803B actually contributed more of the combined light. This could have resulted in an apparently shallower eclipse, which we judge to be an artefact. There are also some low-amplitude $\delta$~Sct pulsations visible in the FT of the S36 data in the frequency range $17.7 - 32$\,d$^{-1}$. These very probably arise from the marginal Am star HD~99803B, which lies within the $\delta$~Sct instability strip. Many marginal Am stars are known to pulsate \citep{2017MNRAS.465.2662S}.These data do not warrant further analysis at present; HD~99803 (HR~4423) will be observed again during TESS S63at, at that time the eclipse depth and p-mode pulsations can be examined further. 

\subsubsection{Eclipses Observed with TESS}

The TESS Sector 10 eclipse occurred at BJD = $2,458,580.6328 \pm 0.0002$, while a second eclipse was detected in Sector 36 at BJD = $2,459,285.8792 \pm 0.0006$.  There are no TESS observations in between these two sectors, which are separated by two years.  The difference between these two times of eclipse is $705.246 \pm 0.001$ days.  Therefore in the absence of any perturbing effects on the binary orbit, the orbital period is $P_{\rm orb} = 705.246/N \pm 0.001/N$ days.

The eclipse depths are about 4\% deep.  For an equatorial occultation by the companion star, the inferred size of the companion star is approximately 20\% of the radius of HD~99803A.  Given that the latter is about 3\,R$_\odot$, this implies a companion radius of roughly 0.6\,R$_\odot$. If so, this close companion would be orders of magnitude fainter than HD~99803A, and would not be expected to contribute noticeably to its spectrum.  There would also be no readily detectable secondary eclipse.

Another consideration is that HD~99803A and B are co-moving Gaia stars where component B is 2.6 magnitudes fainter than component A in $G$, with a 13\,arcsec separation, which is some 2500\,AU on the plane of the sky.  This fainter star is contained within the photometric aperture of HD~99803.  However, if the eclipses were actually hosted by this co-moving companion, then the actual eclipse depth would be close to 50\%.  That is highly implausible.

\subsubsection{Analysis of the MASCARA photometry}


We are also fortunate to have MASCARA coverage of this star over a nearly 6-year interval.  MASCARA (Multi-site All-Sky CAmeRA; \citet{2017A&A...601A..11T}) and bRing \citep{2017A&A...607A..45S} have observed HD~99803 (HR~4423) since early 2017, providing $\sim$163,000 photometric data points, each representing a set of 50 exposures of 6.4\,s. After filtering the data, $\sim$115,000 data points were used for further analysis. MASCARA and bRing have a pixel scale of approximately 1$'$, while aperture photometry has been performed using a 2.5-pixel radius, with local background subtraction in an annulus of $6-21$ pixels radius. The aperture photometry data have been calibrated using \citet{2018A&A...619A.154T} to remove most instrumental artefacts, mainly due to vignetting, PSF variations and inter-pixel variations. Due to the use of fixed cameras, further strong systematics are known to exist at frequencies of 1 cycle per day, 1 cycle per sidereal day (and its harmonics), and synodic and sidereal lunar periods.

\begin{figure}
\begin{center}
\includegraphics[width=0.4\columnwidth,angle=0]{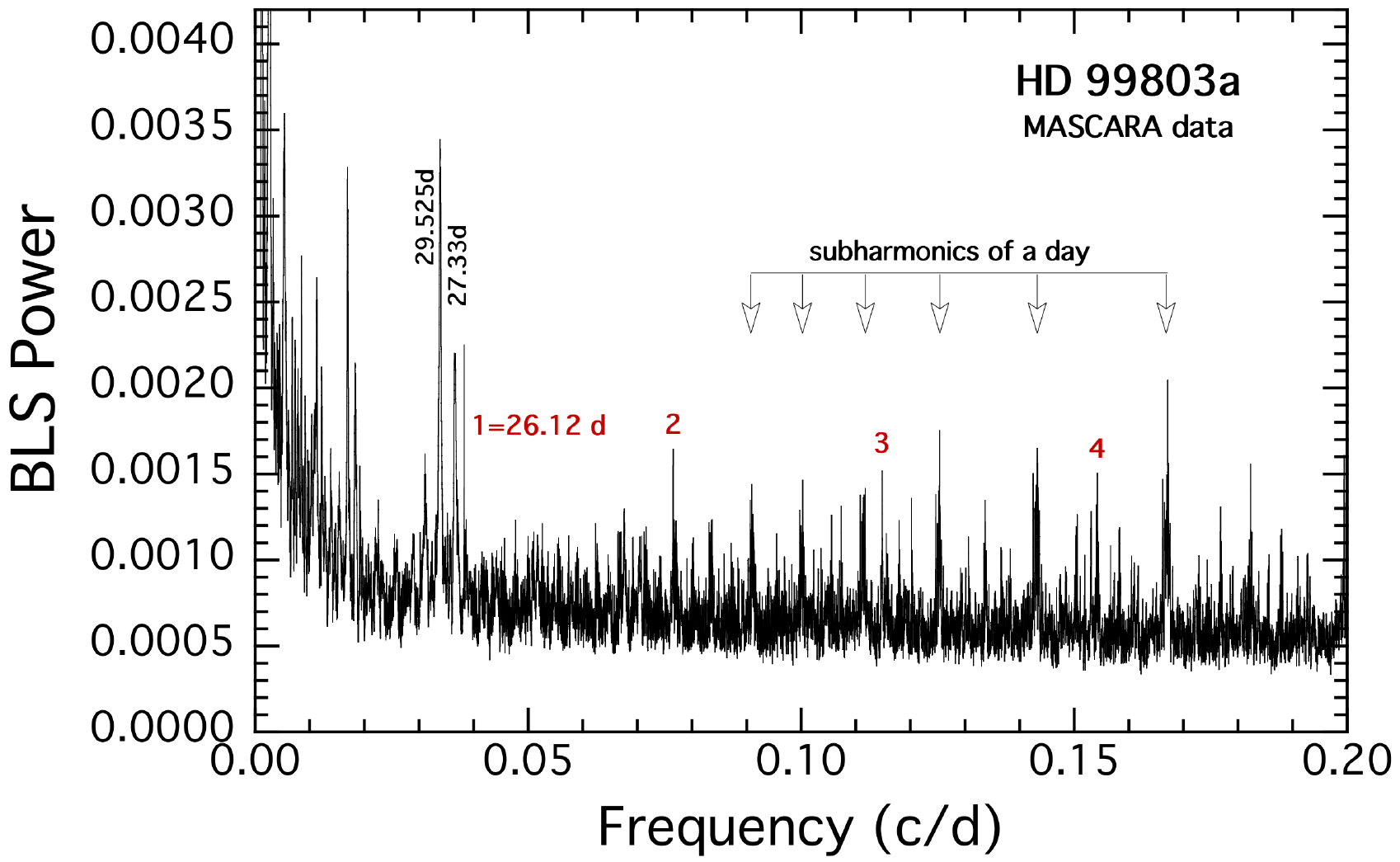}
\includegraphics[width=0.4\columnwidth,angle=0]{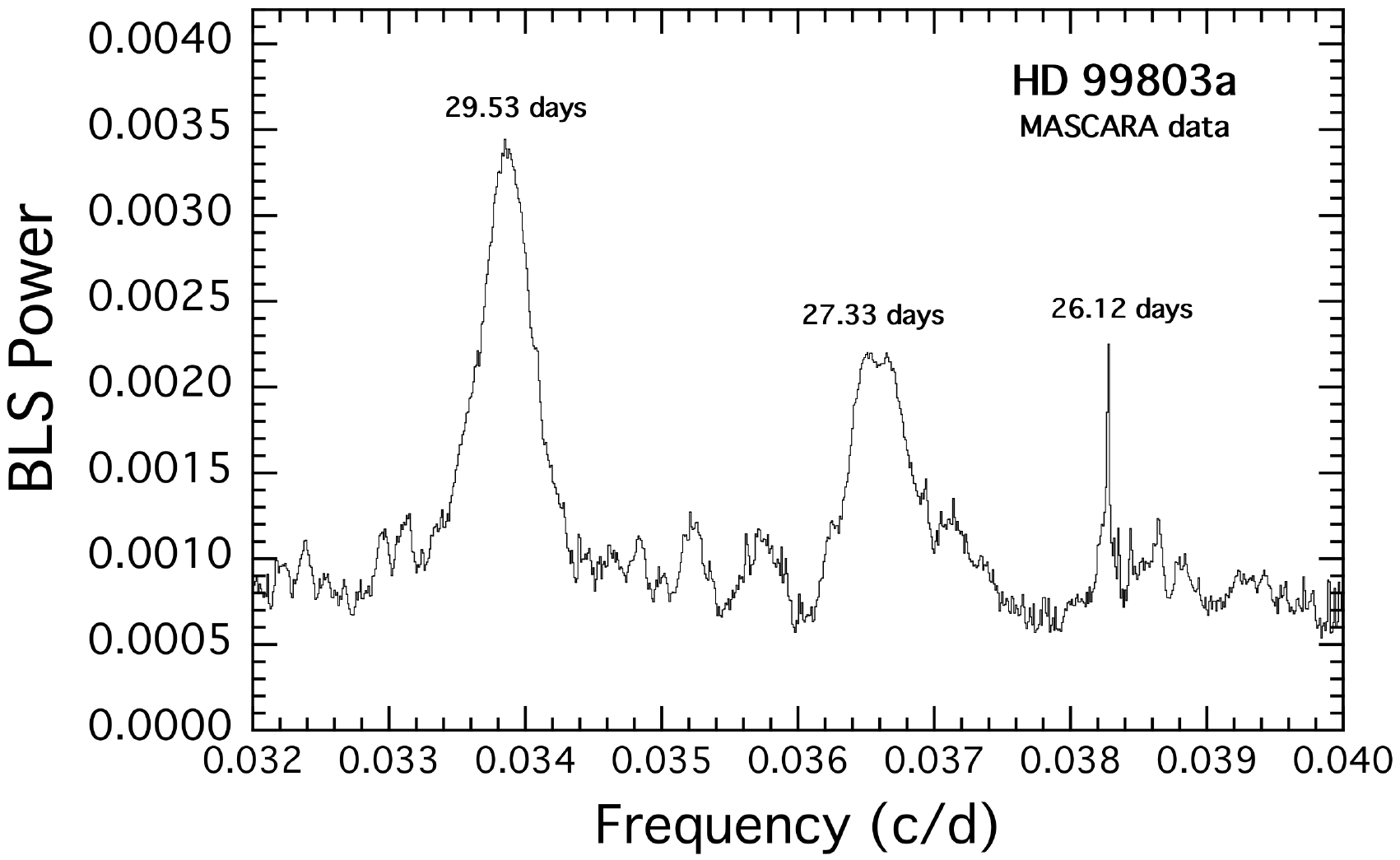}
\includegraphics[width=0.4\columnwidth,angle=0]{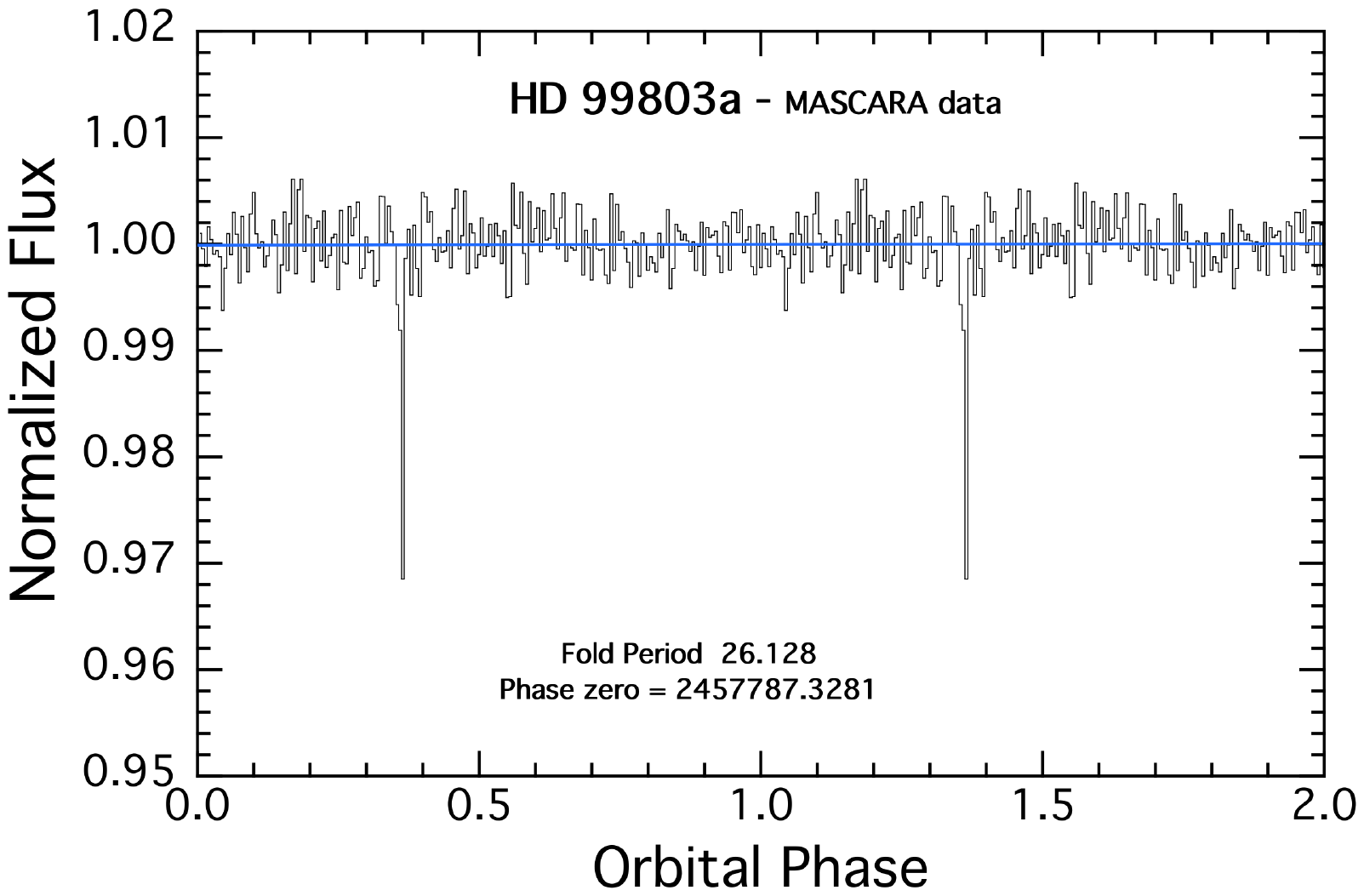}
\caption{Detection of the binary period of HD~99803A with MASCARA data. Upper left panel: BLS transform of 5.4-y of MASCARA photometric data on HD~99803A.  The 26.12-d detected period and its next three higher harmonics are marked in red font.  The synodic and sidereal periods of the Moon are indicated, as are some of the subharmonics of the daily observation cycle. Upper right panel: Zoom-in around the two lunar periods and the eclipse period.  Note that the former are extremely broad, while the eclipse period is appropriately narrow. Bottom panel: Folded, binned, and averaged orbital light curve about a period of 26.128\,d. The eclipse is narrow ($\lesssim 1\%$ of the orbital period) and $\sim$4\% deep, in agreement with the eclipse seen in the Sector 10 TESS data.}
\label{fig:mascara} 
\end{center}
\end{figure}

 The first panel in Fig.~\ref{fig:mascara} shows the Box Least-Squares (BLS) transform of the MASCARA data.  The largest peaks are associated with the synodic and sidereal periods of the lunar cycle which affects the observations.  There are also a number of peaks associated with subharmonics of the daily observing cycle.  In addition to these non-astrophysical peaks, there is one significant peak at a period of 26.128\,d, and its three higher harmonics, labeled 1-4.  The second panel of Fig.~\ref{fig:mascara} shows a zoom-in of the BLS transform around the range of $25-31$\,d. As one can readily see, the peak at 26.128\,d is much narrower than either peak associated with the lunar activity. Furthermore, the four harmonics of this period indicate that it represents a sharp feature in the light curve.  The third panel in this figure shows a fold of the MASCARA data about the 26.128-d period.  It is gratifying to see that it is a narrow eclipse that is about 4\% deep, just as the TESS Sector 10 eclipse indicates.  
 
 Finally, we note that the phase of the MASCARA eclipse agrees with the TESS eclipses to within $\lesssim 1\%$ of an orbital cycle.  The MASCARA period therefore represents the same eclipsing behavior we see in the TESS data.  This, in turn, implies that the `missing' integer $N$ in the expression above for the orbital period is 27. The accurate period determined from the time difference between the two TESS eclipses, and the integer number of cycles between them, is then $26.12022 \pm 0.00004$ days.

\subsection{Frequency analysis of TESS data - HD~1279 (HR~62)} 
\label{sec:fa2}

\begin{figure}
\begin{center}
\includegraphics[width=1.0\linewidth,angle=0]{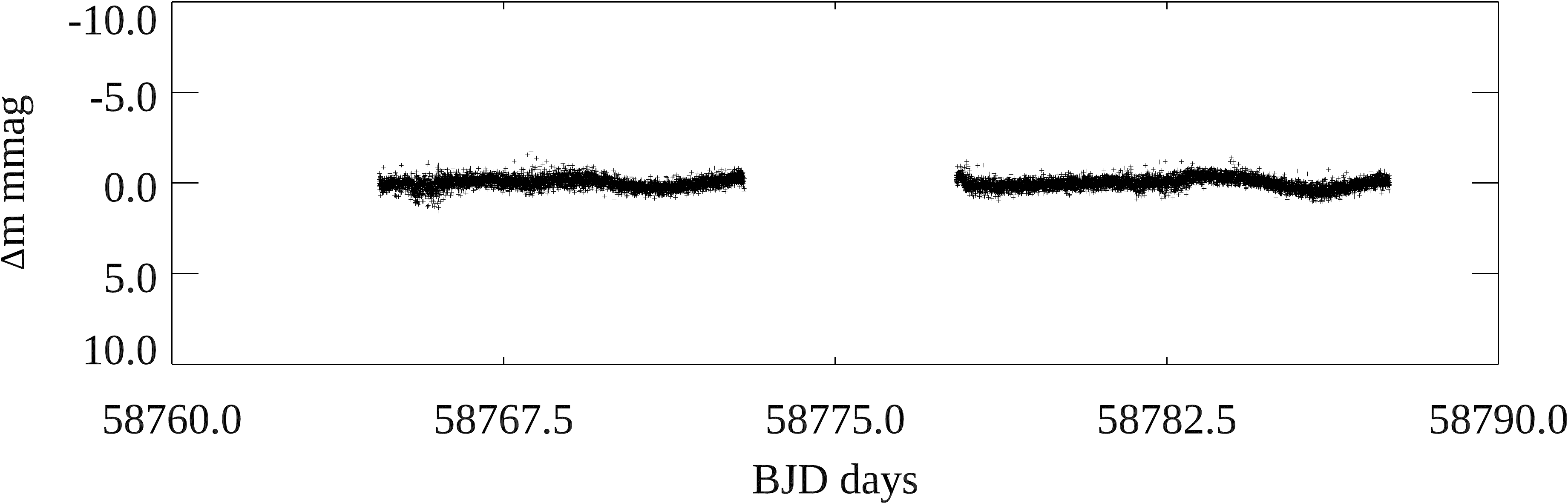}\caption{The light curve of HD~1279 (HR~62) obtained in 120-s cadence in Sector 17. The ordinate scale is Barycentric Julian Date -- 240\,0000.0.  }
\label{fig:lc2} 
\end{center}
\end{figure}

HD~1279 (TIC\,440076466; HR\,62) was observed by {\em TESS} in Sector 17 in 2-min cadence. The data are available in both SAP (simple aperture photometry) and PDCSAP (presearch-data conditioning SAP); we used the PDCSAP data for our analysis, converting intensity to magnitudes. The data have a time span of 22.85\,d with a centre point in time of $t_0 = {\rm BJD}~2458776.11438$, and comprise 12969 data points, after clipping 9 outliers, which are probably the result of cosmic ray strikes (including solar radiation).  The TESS lightcurve od HD~1279 (HR~62) is shown in Fig.\,\ref{fig:lc2}.

\begin{figure}
\begin{center}
\includegraphics[width=1.0\linewidth,angle=0]{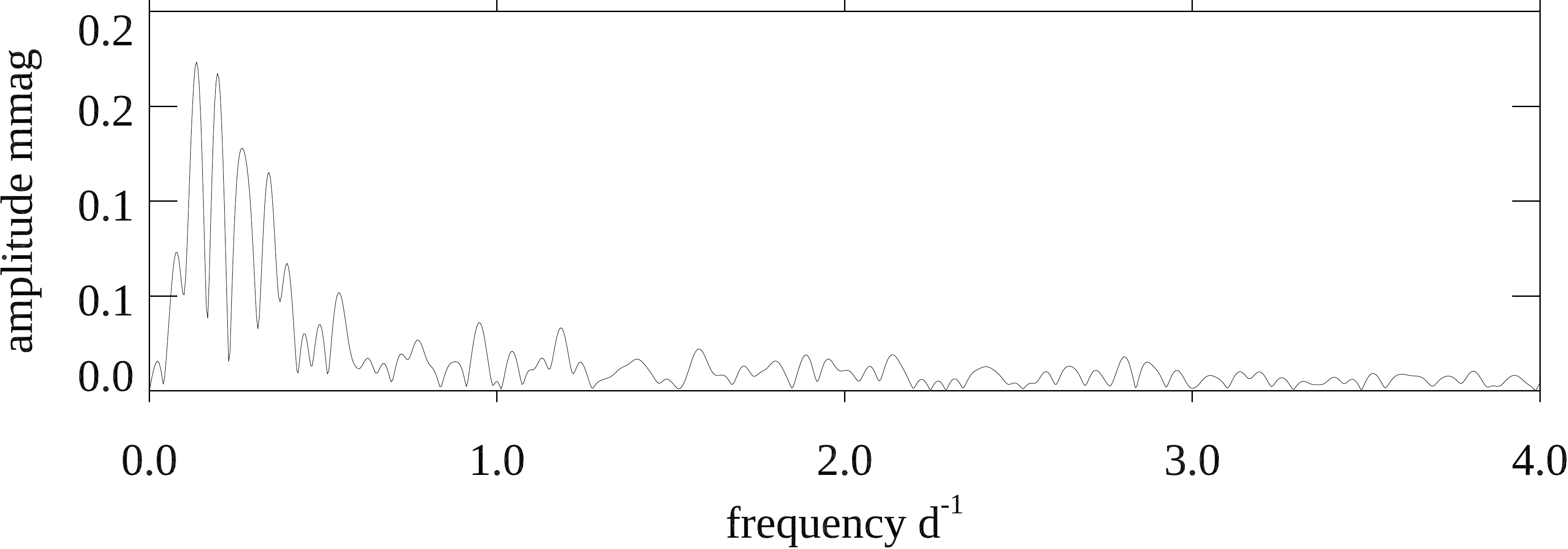}
\caption{An amplitude spectrum of the Sector 17 data for HD~1279 (HR~62). There is no evidence of any pulsation out to the Nyquist frequency near 360\,d$^{-1}$. This plot shows on the low frequency range where there is a possibility of g~mode frequencies in a B8III star. No peaks that cannot be ascribed to instrumental variations are seen.}
\label{fig:ft2} 
\end{center}
\end{figure}

The S17 data were analysed using a fast Discrete Fourier Transform \citep{1985MNRAS.213..773K} to produce an amplitude spectrum (Fig.\,\ref{fig:ft2}).  There is no evidence of any p~mode pulsations out to the Nyquist frequency near 360\,d$^{-1}$ (not shown). In the low frequency range where g~modes occur there is no peak that cannot be ascribed to instrumental variations. Hence there is also no evidence of g~modes. Comparison with the instability strips of \citet{2008JPhCS.118a2079Z} show that HD~1279 is too cool to be a $\beta$\,Cep star and too luminous to be an SPB star.

\subsection{Frequency analysis of TESS data - HD~123445 (HR~5294)}
\label{sec:fa3}

\begin{figure}
\begin{center}
\includegraphics[width=1.0\linewidth,angle=0]{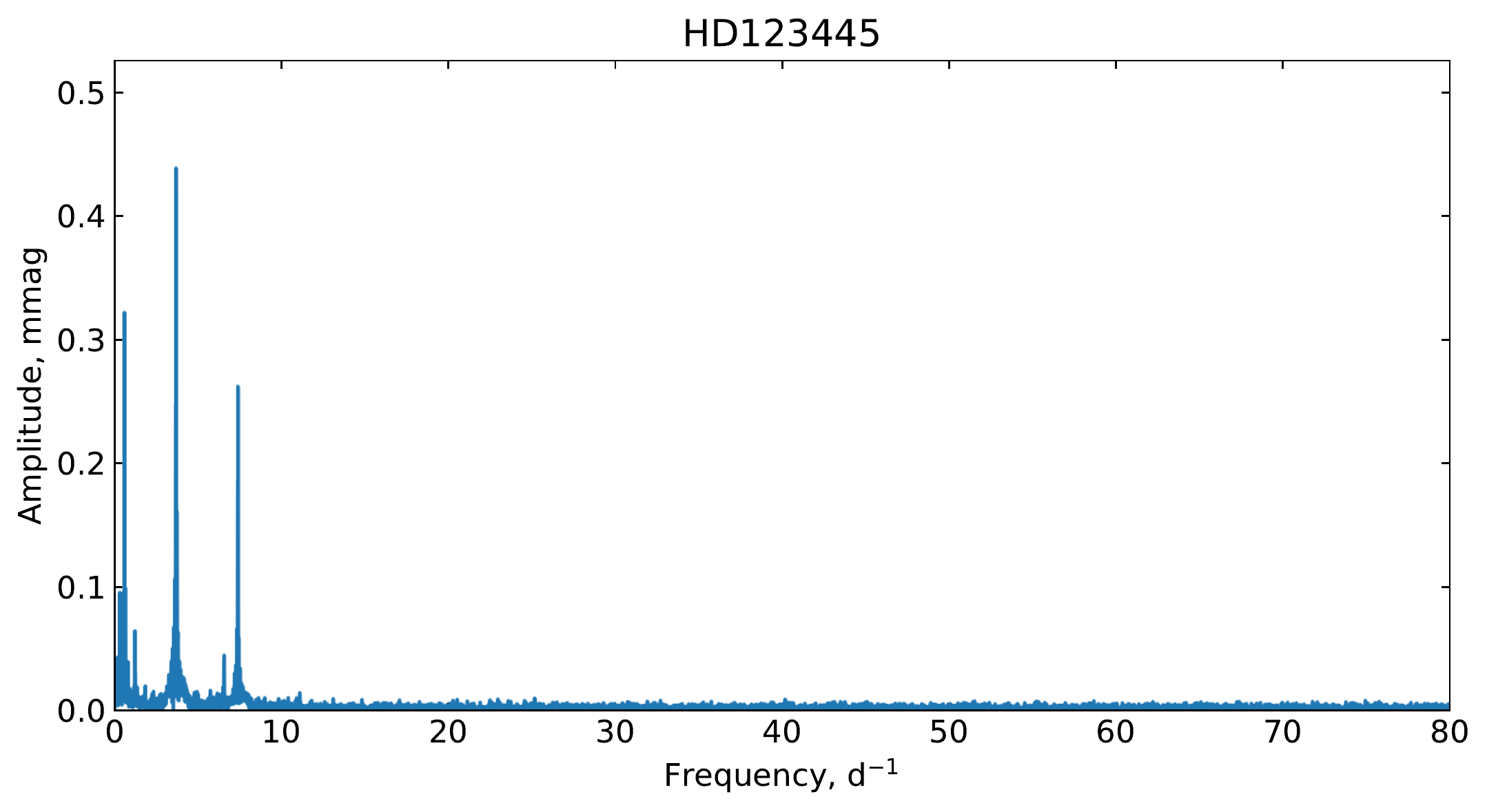}
\caption{An amplitude spectrum of the Sector 17 data for HD~123445 (HR~5294). }
\label{fig:ft3} 
\end{center}
\end{figure}

HD~123445 (TIC\,242288894; HR\,5294) was observed by {\em TESS} in Sector 11 in 30-min cadence. 
The Fourier Transform of the light curve from sector 38 (2-min cadence) reveals that HD~123445 (HR~5294) has a rotation period of roughly 3 days. It also pulsates with a frequency of about $f_{1}$ = 3.8 c/d, and the harmonics  2 $f_{1}$ and $f_{1}$ show up clearly (Fig.\,\ref{fig:ft3}). The frequencies are collected in Table \ref{tab:HD123445}

\begin{table}
\centering
\caption{The frequencies of HD~123445 (HR~5294)}
\begin{tabular}{cccc}
\hline
Designation & 	Frequency d$^{-1}$ & Amplitude (mag) & Comment \\
\hline
F5 	& 0.35658876 	& 0.000083  &       \\	
F4 	& 0.62403082 	& 0.000319  &      \\	
F6 	& 1.25185483 	& 0.000059  &  \\	
F1 	& 3.71004238 	& 0.000448  &   \\	
F7 	& 3.75556579 	& 0.000149  &	 \\ 	
F8 	& 6.59310162 	& 0.000043  &  \\	
F2 	& 7.42008475 	& 0.000260  & 2f1   \\
F3 	& 11.1301271 	& 0.000013  & 3f1   \\
\hline
\label{tab:HD123445}			      		     	  
\end{tabular}
\end{table}

\subsection{Frequency analysis - HD~147550 (HR~6096), HD~171961 (HR~6990) and HD~202671 (HR~8137)}
\label{sec:fa4}

\textbf{HD~147550 (TIC\,71983021; HR\,6096),  HD~171961 (TIC\,1286643; HR\,6990) and HD~202671 (TIC\,39364873; HR\,8137) were not observed by {\em TESS}.}

\section{Model atmosphere analysis and synthetic spectra computation}

For the six stars rotating with a \vsini  lower than 30 \kms, the abundances or their upper limits for forty chemical elements could be derived by iteratively
adjusting synthetic spectra to the normalized spectra and looking for the best fit to carefully selected unblended lines.
For the two stars rotating faster than 30 \kms, HD~123445 (HR~5294) and HD~171961 (HR~6990), the lines of elements with intrinsically low abundances (eg. the lanthanides and heavy elements) are blended and fewer abundances could be derived.
Specifically, synthetic spectra were computed assuming
LTE using \cite{Hubeny} SYNSPEC49 code which computes line profiles for
elements up to Z=99.

\subsubsection{Microturbulent velocity determinations}

\begin{figure}[h!]
\vskip 0.5cm
   \centering
      \includegraphics[scale=0.90]{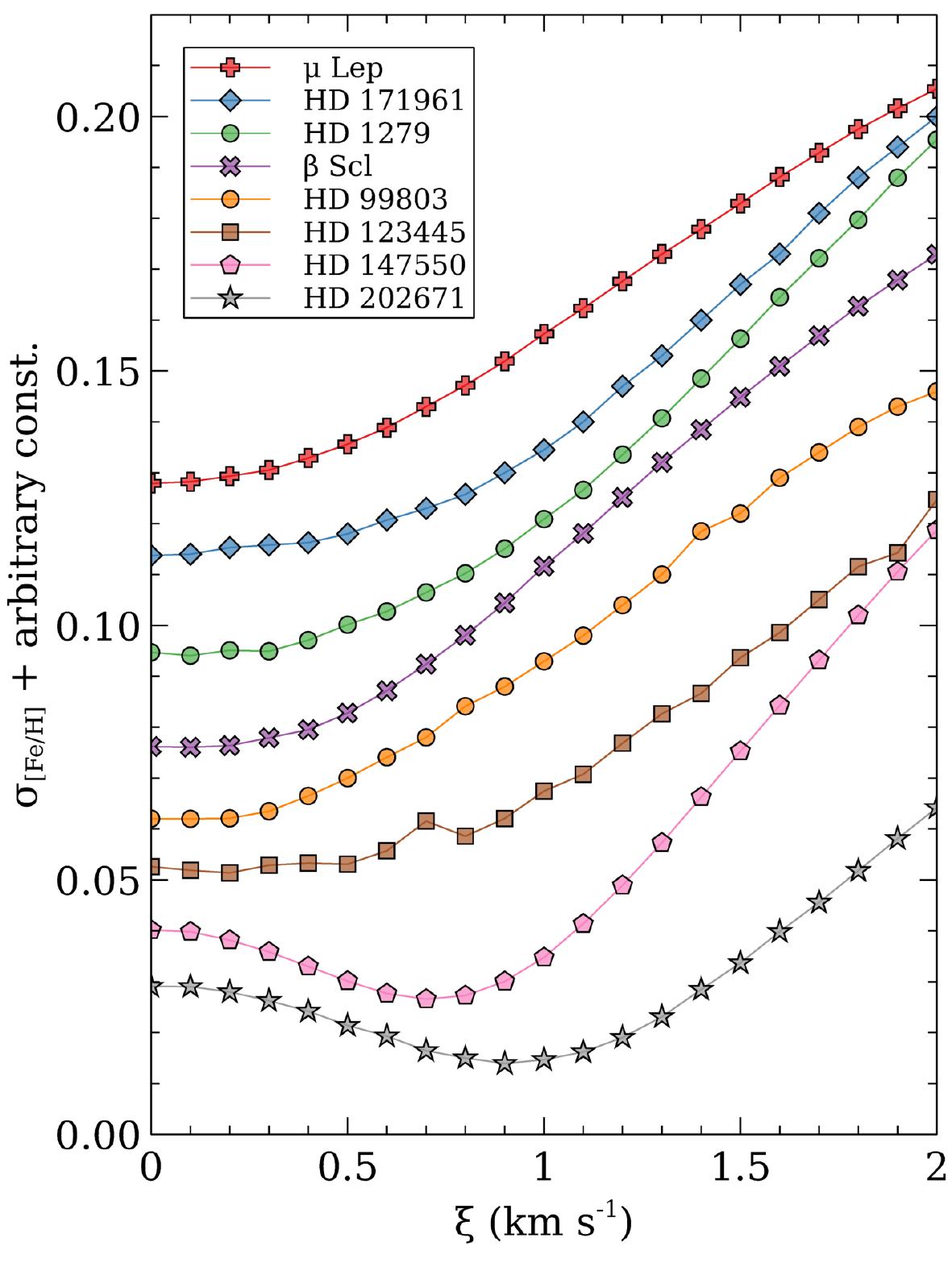}
   \caption{Microturbulent velocity determinations for the eight stars
   \label{figure2}}\vskip 0.5cm
\end{figure}

In order to derive the microturbulent velocities, we have derived the iron abundance
[Fe/H] from unblended \ion{Fe}{2} lines for a set of microturbulent
velocities ranging from 0.0 to 2.0 \kms for each star. Figure \,\ref{figure2} shows the standard
deviation of [Fe/H] as a function of the microturbulent velocity.
The adopted microturbulent velocities are the values which minimize the standard deviations, i.e., for that value, all \ion{Fe}{2} lines yield the same iron abundance.
The microturbulent velocities of the eight stars, which are accurate to within $\pm$ 0.1 \kms, are collected in Table \ref{tab:fundamentals}. Only two stars, HD~147550 (HR~6096) and  HD~202671 (HR~8137), have microturbulent velocities which differ from 0 \kms.


\subsection{Model atmospheres}

The ATLAS9 code \citep{Kurucz} was used to compute a first model atmosphere for the effective temperature and surface gravity of each star assuming
a plane parallel geometry, a gas in hydrostatic and radiative equilibrium and local thermodynamical equilibrium. The ATLAS9 model atmosphere contains 72 layers with a regular increase in
$\log \tau_{Ross} = 0.125$ and was calculated assuming a solar chemical composition \citep{1998SSRv...85..161G}. It was converged up to $\log \tau = -5.00$ in order to attempt reproduce the cores of the Balmer lines.
This ATLAS9 version uses the new opacity
distribution function (ODF) of \cite{Castelli2003} computed for that solar
chemical composition.
Once a first set of elemental abundances was derived using the ATLAS9 model atmosphere, the atmospheric structure was recomputed for these abundances using the Opacity Sampling ATLAS12 code \citep{atlas12,atlas12_2}.
New slightly different abundances were then derived and a new ATLAS12 model recomputed until the abundances of iteration (n-1) differed of those of iteration (n) by less than $\pm$ 0.10 dex.

\subsection{The linelist}

Atomic linelists from Kurucz's database have provided the basis to construct our linelist\footnote{http://kurucz.harvard.edu/lineslists/ }. These lists collect data mostly from the literature for light and heavy elements (usually critically evaluated transition probabilities from \cite{Martin88} and \cite{Fuhr1988}) and computed data by \cite{Kurucz} for the iron group elements.
A first linelist was built from Kurucz's file gfall21oct16.dat \footnote{\url{http://kurucz.harvard.edu/linelists/gfnew/gfall21oct16.dat}} which includes hyperfine splitting levels.
We then updated several oscillator strengths and damping parameters with more recent and accurate determinations when necessary.
As a rule, we have preferred NIST\footnote{http://physics.nist.gov/cgi-bin/AtData/linesform} and \cite{W96} oscillator strengths for CNO and also \cite{2006A&A...445.1165N}. 
The H I lines are calculated using \cite{1973ApJS...25...37V} tables upgraded by 
Schoening \& Butler up to H 10.
The He I lines are computed in SYNSPEC49 by using 
specific tables, either from \cite{shamey} or \cite{Dimitrijevic1984}.
%

We introduced hyperfine components for a few lines of Mn II \citep{1999MNRAS.306..107H}, isotopic and hyperfine transitions for the \ion{Ga}{2} line at 6334.07 \AA\  \citep{2000A&A...363..815N}.
To model the \ion{Hg}{2} line at 3983.93 \AA, we have included 9 hyperfine transitions from all isotopes from $\mathrm{Hg}^{196}$ up to $\mathrm{Hg}^{204}$ from \cite{Do03}. 
\textbf{For \ion{Pt}{2}, we have used the atomic data provided in \citep{1989ApJ...340.1140E,1998ApJ...504..533B,1998CoSka..27..324J}}.

For the Rare Earths, we retrieved all relevant transitions from the DREAM database\footnote{http://www.umh.ac.be/astro/dream.shtml}.
We also used specific publications reporting on laboratory measurements: Sm II \citep{LA06}, Nd II \citep{2003ApJS..148..543D}, Eu II \citep{LWHS}, Tb II \citep{law01}.
In Kurucz's linelists, the damping constants are taken from the literature when available. For the iron group elements, they come from Kurucz's (1992) computations. Additional damping \textbf{constants} for a few Si II lines were taken from \cite{1988A&A...192..249L}. When they are not available from the linelist, damping constants are actually computed in SYNSPEC49 using an approximate formula \citep{1981SAOSR.391.....K}.

\subsection{Spectrum synthesis}

We have used only unblended lines to derive the abundances. For the two fastest rotators, HD~123445 (HR~5294) and HD~171961 (HR~6990), fewer abundances could be derived.
For a given element, the final abundance is a weighted mean of the abundances derived for each transition. The weights are derived from the NIST grade assigned to that particular transition when available. The same weight was assigned to all transitions when no critical evaluation from NIST was available. 
 
For several elements, in particular the heaviest elements, only one unblended line was available and the abundance should be regarded as an estimate only.
For each modeled transition, the adopted abundance is that which provides the best fit to the observed normalized profile.
A grid of synthetic spectra was computed with SYNSPEC49 \citep{Hubeny} to model each selected unblended
 line of the forty elements. Computations were iterated varying the unknown abundance until minimization of the chi-square between the observed and synthetic spectrum was achieved.

\section{Abundance determinations}

We discuss here the identifications of chemical elements, the lines selected for abundance determinations  and the abundance determinations for each element.
 We also provide non-LTE abundance corrections for three light
elements, namely \ion{Na}{1}, \ion{Mg}{1} and \ion{Ca}{2}. 
We use the Formato code \citep{2011MNRAS.418..863M} to build simple model
atoms, using atomic data from NIST \citep{Kramida2018} for energy
levels, from VALD \citep{rya2015} for radiative bound-bound
transitions and from TOPbase \citep{1992RMxAA..23..107C} for radiative
bound-free transitions. For inelastic collisions with electrons, we use
the semi-empirical formula from \cite{seaton62} for which a radiative
counterpart exist and a collision strength of one for transitions with a
forbidden radiative couterpart. For inelastic collisions with hydrogen,
we use the formula from \cite{drawin1969} without scaling factor, except
for \ion{Mg}{1} where we implemented the mechanical quantum data from
\cite{GUITOU201594}.
The statistical equilibrium and the radiative
transfer equation for each level and line in each model atom are
consistently solved using the non-LTE 1D radiative transfer code MULTI
\citep{Carlsson86,Carlsson92}.

\subsection{Helium}

The strongest helium lines unambiguously detected are: $\lambda$  4026.19 \AA, $\lambda$ 4471.48 \AA, $\lambda$ 5875.621 \AA. The others are either weak or embedded into blends.
The least blended, $\lambda$ 4471.48 \AA, has been synthesized to derive the helium LTE abundance. Helium is underabundant in all stars, the deficiency is the most important in HD~171961 (HR~6990), for which the abundance is only 4 \% of the solar abundance \textbf{(-1.07)}




\subsection{The light elements (C to Ti)}

\subsubsection{Carbon}

Most of the expected \ion{C}{2} lines are blended with more abundant species. The least blended is the high excitation \ion{C}{2} triplet at 4267.26 \AA.
Carbon is underabundant in HD~1279 (HR~62), HD~99803 (HR~4423), HD~123445 (HR~5294), HD~147550 (HR~6096), HD~202671 (HR~8137), $\beta$ Scl (HR~8937) and $\mu$ Lep (HR~1702) with similar deficiencies, from 0.30 up to 0.60 times the solar abundance \text{(-1.07)}.

Carbon is very underabundant in HD~171961 (HR~6990). \cite{1987SvA....31..151S} has shown that the \ion{C}{2} line at 4267.26 \AA\ is prone to NLTE effects for B stars hotter than 15000 K. He suggests to use the \ion{C}{2} lines at 3919.97 and 3921.69 \AA\ less influenced by departures from LTE to derive carbon abundances. These lines yield the same carbon abundances as 4267.26 \AA\ for the eight stars.

\subsubsection{Oxygen}

For \ion{O}{1}, the strongest expected allowed lines are those of multiplet 7 (NIST quality B+). The nine transitions of multiplet 7 dominate the opacity from 6155 \AA\ to 6159 \AA. 
Oxygen is underabundant in HD~99803 (HR~4423), HD~123445 (HR~5294), HD~147550 (HR~6096), HD~171961 (HR~6990), HD~202671 (HR~8137) and $\beta$ Scl (HR~8937) with depletions of about 0.50 to 0.80 times the solar abundance \textbf{(-3.31)}. In $\mu$ Lep (HR~1702), oxygen is about solar and marginally overabundant in HD 1279 (HR~62).

\subsubsection{Neon}

The \ion{Ne}{1} line at 6402.25 \AA\ (NIST quality AAA) was used to derive neon abundances. Neon is overabundant in HD~202671 (HR~8137) and quite underabundant in $\beta$ Scl (HR~8937) and $\mu$ Lep (HR~1702). In all other stars, $\lambda$ 6402.25 \AA\ is not observed which indicates an upper limit of one solar abundance \textbf{(-4.07)}. 

\subsubsection{Sodium}

We have used the 2 \ion{Na}{1} lines at 4497.66 \AA\ (NIST quality B+) and the resonance \ion{Na}{1} lines at 5889.95 and 5895.92 \AA\ (quality AA). 
The LTE overabundances of sodium derived from these lines are \textbf{-5.67 (ie. solar)} for HD 1279 (HR~62) and HD 171961 (HR~6990), half the solar value for HD~99803 (HR~4423), and overabundant by a factor of 4 in $\beta$ Scl (HR~8937).
No conclusion could be reached for HD 123445 (HR~5294) because of telluric lines and for HD 202671 (HR~8137) and $\mu$ Lep (HR~1702) because of blending.

The NLTE corrections for the resonance \ion{Na}{1} lines at 5889.95 and 5895.92 \AA\ are about -0.52 and -0.60 dex respectively which yields Non-LTE abundances of Na of 
-6.27 in HD~1279 (HR~62) and HD~171961 (HR~6990),  -5.97 for HD~99803A (HR~4423) and -5.37 for $\beta$ Scl (HR~8937), well below the LTE determinations.

\subsubsection{Magnesium}

The unblended \ion{Mg}{2} lines at  4390.514 \AA\ and 4390.572 \AA\ yield LTE abundances of -4.94 (HD 1279, HR~62), -5.0 (HD~99803A, HR~4423) , -4.72 (HD 123445 (HR~5294), HD 171961 (HR~6990), $\beta$ Scl (HR~8937)),
-4.57 (HD 147550, HR~6096), -4.77 (HD 202671, HR~8137), -4.54 ($\mu$ Lep, HR~1702) respectively \textbf{the solar abundance is -4.42}.
The unblended \ion{Mg}{1} line at 5172.68 \AA\ yields similar abundances, after a NLTE correction which amounts to -0.33 dex. 
Magnesium therefore appears to be underabundant in each star.

\subsubsection{Aluminium}

The synthesis of the unblended  \ion{Al}{2} line at 4663.056 \AA\ shows that aluminium is severely depleted in all stars except in HD 147550 (HR~6096) where it is overabundant (-5.23).
The underabundance is the most pronounced in $\beta$ Scl (HR~8937, -6.83). \textbf{The solar abundance is -5.53}.

\subsubsection{Silicon}

The synthesis of strong and unblended lines of \ion{Si}{2} reveals a large spread of the silicon abundances: from very deficient (-5.75) in HD 1279 (HR~62) to overabundant in HD 147550 (HR~ 6096,-4.25), $\beta$ Scl (HR~8937, -4.09) and $\mu$ Lep (HR~1702, -3.91). \textbf{The solar silicon abundance is -4.45}.

\subsubsection{Phosphorus}

The lines of \ion{P}{2} expected strongest are $\lambda$ 6024.18 \AA\ and 6043.12 \AA\ of multiplet 2. Phosphorus is overabundant in HD 1279 (HR~62, -5.07), HD 147550 (HR~6096, -5.67), HD 171961 (HR~6990, -5.00), HD 202671 (HR~8137, -4.32), $\mu$ Lep (HR~1702, -5.27) and $\beta$ Scl (HR~8937, -5.14).
None of the lines mentioned above are detected in HD~99803A (HR~4423) and HD 123445 (HR~5294), suggested a 1 solar upper limit. \textbf{The solar phosphorus abundance is -6.59}.

\subsubsection{Sulfur}
 
 The lines of \ion{S}{2} at 4162.67 \AA, 4142.25 \AA\ and 4153.06 \AA\ of multiplet 4 are clearly present and unblended in the spectra of all stars. 
Sulfur is solar in HD~123445 (HR~5294), HD~147550 (HR~6096) and $\mu$ Lep (HR~1702, -4.88) and underabundant in all other stars. \textbf{The solar sulfur abundance is -4.67}.

\subsubsection{Argon}
 
The high excitation line of \ion{Ar}{2} at 5194.53 \AA\ was used to derive the abundance of argon.
Argon is solar in HD~1279 (HR~62, -5.60). The \ion{Ar}{2} line is blended in HD~202671 (HR~8137).
The line of \ion{Ar}{2} is not detected in HD~99803 (HR~4423), HD~123445 (HR~5294), HD~147550 (HR~6096), hD~171961 (HR~6990), $\mu$ Lep (HR~1702) and $\beta$ Scl (HR~8937), which suggests a 1 solar upper limit. \textbf{The solar argon abundance is -5.82}.

\subsubsection{Calcium}

The synthesis of the unblended \ion{Ca}{2} lines at 3933.66 \AA\ and 5019.97 \AA\ yields the calcium LTE abundances, respectively
-5.64 (HD~1279, HR~62), -6.34 (HD~99803A, HR~4423), -5.86 (HD~123445, HR~5294), -5.86 (HD~147550, HR~6096), -6.64 (HD~171961, HR~6990), -5.34 (HD~202671, HR~8137), -5.21 ($\mu$ Lep, HR~1702), -5.24 ($\beta$ Scl, HR~8937). \textbf{The solar calcium abundance is -5.64}.
The Non-LTE correction for the 3933.66 \AA\ amounts to +0.37 dex,  Calcium therefore appears to be overabundant in HD~1279 (HR~62), HD~202671 (HR~8137), $\mu$ Lep (HR~1702) and $\beta$ Scl (HR~8937).

\subsubsection{Scandium}

The line of multiplet 1 of \ion{Sc}{2} at 4246.820 \AA\ is the only unblended line (quality A' in NIST). Scandium appears to be overabundant in HD~1239 (HR~62, -8.13),
HD~99803A (HR~4423, -8.35) and $\mu$ Lep (HR~1702, -8.13) and underabundant in HD~147550 (HR~6096, -9.13) and $\beta$ Scl (HR~8937, -9.83). The line is not detected in HD~123445 (HR~5294), HD~171961 (HR~6990) and HD~202671 (HR~8137), for which we adopt a 1 solar upper limit (\textbf{the solar scandium abundance is -8.85}).

\subsubsection{Titanium}

Ten lines of \ion{Ti}{2} were used to derive the mean titanium abundances. Titanium is found to be overabundant in all stars, from marginally overabundant in HD~147550 (HR~6096, -6.82) to a sizeable overabundance (-6.10) in $\beta$ Scl (HR~8937). \textbf{The solar titanium abundance is -6.98}.

\subsection{The iron-peak elements (V to Zn) and gallium}

\subsubsection{Vanadium}

The vanadium abundance has been estimated using three lines of \ion{V}{2} at  4005.705 \AA, 4023.378 \AA\ and 4036.78 \AA. 

The \textbf{analysis} of these lines yields an overabundance in HD~147550 (HR~6096), a solar abundance in HD~1239 (HR~62), HD~123445 (HR~5294), HD~171961 (HR~6990) and $\beta$ Scl (HR~ 8937, -8.07) and a one solar upper limit for HD~202671 (HR~8137) and $\mu$ Lep (HR~1702). \textbf{The solar vanadium abundance is -8.00}. The selected lines of \ion{V}{2} are not observed in these two stars.

\subsubsection{Chromium}

Several strong and unblended lines of \ion{Cr}{2} could easily be found in the spectra of all tars. Six unblended lines of \ion{Cr}{2}  were synthesized. 
Chromium is solar in HD~171961 (HR~6990, -6.36) and overabundant in all stars from mildly overabundant in HD~1239 (HR~62), HD~123445 (HR~5294), HD~147550 (HR~6096) and HD~202671 (HR~8137) to overabundances close to 5 times the solar chromium abundance in HD~99803A (HR~4423), $\mu$ Lep (HR~1702) and $\beta$ Scl (HR 8937). \textbf{The solar chromium abundance is -6.33}.

\subsubsection{Manganese}

The two \ion{Mn}{2} lines have hyperfine structures published by \cite{1999MNRAS.306..107H}. They yield overabundances in all stars, but for HD~171961 (HR~6990) which has a solar manganese abundance (-6.57). The overabundances are very pronounced in HD~1279 (HR~62, -5.21), HD~202671 (HR~8137, -5.10), $\mu$ Lep (HR~1702, -4.13) and $\beta$ Scl (HR~8937, -4.21) and mild in HD~99803A (HR~4423, -6.00), HD~123445 (HR~5294, -6.21), HD~147550 (HR~6096, -6.21). \textbf{The solar manganese abundance is -6.61}.
\\

\subsubsection{Iron}

The abundance of iron was derived from eleven \ion{Fe}{2} lines, all of which are sensitive to small changes in $[\frac{Fe}{H}]$. 
Iron is mildly overabundant in HD~202671 (HR~8137, -4.10), about solar in HD~1279 (HR~62), HD~147550 (HR~6096), HR~6990, $\mu$ Lep (HR~1702) and $\beta$ Scl (HR~8937, -4.50) and very underabundant in HD~99803A (HR~4423, -4.90) and HD~147550 (HR~6096, -5.02). \textbf{The solar iron abundance is -4.50}.

\subsubsection{Nickel}

The nickel abundances was derived using one line at 4067.031 \AA\ only and should be taken with caution. Nickel is found to be underabundant in HD~1279 (HR~62), HD~99803A (HR~4423), HD~171961 (HR~6990), HD~202671 (HR~8137) and $\beta$ Scl (HR~8937, --6.40 for these stars), very underabundant in $\mu$ Lep (HR~1702, -7.16) and HD~123445 (HR~5294, -7.00), and close to solar in HD~147550 (HR~6096, -5.78). \textbf{The solar nickel abundance is -5.75}.

\subsubsection{Gallium}

We have used the high excitation \ion{Ga}{2} line at $\lambda$ 6334.069 \AA\ with hyperfine splitting of the gallium isotopes. It is the strongest expected line of \ion{Ga}{2} in this temperature regime. The eight transitions retrieved and modified from 
\cite{2000A&A...363..815N} are collected in Table \ref{tab:Gaii}.
\\
The line of \ion{Ga}{2} at 6334.069 \AA\ is usually clearly present. Other lines of \ion{Ga}{2} are present at 4251.108 \AA, 4254.032 \AA, 4255.64 \AA\ and 4261.995 \AA\ and 5360.313 \AA, 5363.353 \AA\ and 5421.122 \AA\ in the red part of the spectrum. 
Gallium is overabundant in HD~1279 (HR~62, -7.18), HD~123445 (HR~5294, -5.96), HD~171961 (HR~6990, -6.86), $\mu$ Lep (HR~1702, -5.38), $\beta$ Scl (HR~8937, -5.96),
solar in HD~99803A (HR~4423), HD~147550 (HR~6096, -8.96), consistent with an upper limit of one solar in HD~202671 (HR~8137, \textbf{the solar gallium abundance is -8.96}, no line observed).

\begin{table}
\centering
\caption{The \ion{Ga}{2} $\lambda$ 6334.069 \AA\ linelist}
\begin{tabular}{ccccc}
\hline
Wavelength & 	ion 	   & $\log gf$ & lower energy level & EW \\
(\AA) & & & (cm$^{-1}$) & (m\AA) \\ \hline
  6333.930 &  \ion{Ga}{2}  &  0.08 & 102944.595 &   14.7 \\
  6333.980 &  \ion{Ga}{2}  &  0.21 & 102944.595 & 20.5  \\
  6333.990 &  \ion{Ga}{2}  &  0.08 & 102944.595 & 14.6  \\
  6334.069 & \ion{Ga}{2}   &  0.10 & 102944.595 & 15.3 \\
  6334.080 &  \ion{Ga}{2}  &  0.40 & 102944.595 & 26.5  \\
  6334.083 &  \ion{Ti}{2}  & -2.08 &  66521.008 & 0.0   \\
  6334.120 &  \ion{Ga}{2}  &  0.02 & 102944.595 & 12.8 \\
  6334.200 &  \ion{Ga}{2}  &  0.09 & 102944.595 & 20.9  \\
  6334.208 &  \ion{Cr}{2}  & -1.66 &  89812.422 & 0.0  \\
  6334.290 &  \ion{Ga}{2}  &  0.01 & 102944.595 & 16.8  \\
  6334.368 &  \ion{Ni}{2}  & -2.36 & 120271.975 & 0.0  \\ \hline
\label{tab:Gaii}			      		     	  
\end{tabular}
\end{table} 



%

\subsection{The Sr-Y-Zr triad}

\subsubsection{Strontium}

The abundance of strontium was derived from the fit to the only unblended line of \ion{Sr}{2} at 4215.52 \AA. Strontium is overabundant in HD~1279 (HR~62, -7.35), HD~147550 (HR~6096, -8.13),
$\mu$ Lep (HR~1702, -6.43), $\beta$ Scl (HR~8937, -5.95), solar in HD~99803A (HR~4423, -9.13), HD~123445 (HR~5294), HD~202671 (HR~8137) and HD~171961 (HR~6990). \textbf{The solar strontium abundance is -9.03}.

\subsubsection{Yttrium}

The yttrium lines are conspicuous in the spectra of HD~1279 (HR~62), HD~99803A (HR~4423), HD~123445 (HR~5294), HD~147550 (HR~6096), HD~202671 (HR~8137), $\mu$ Lep (HR~1702), $\beta$ Scl (HR~8937) and HD~147550 (HR~6096). These lines absorb from a few \% up to 10 \%.
The only unblended lines are the low excitation lines at 3982.592 \AA\ and 5662.925 \AA. Yttrium is very overabundant in all stars from a factor of 5 times (- 9.09) up to 12500 times (-5.69) the solar abundance except in HD~171961 (HR~6990) where it is solar (\textbf{the solar yttrium abundance is -9.79}).

\subsubsection{Zirconium}

The  only two \ion{Zr}{2} unblended lines are $\lambda$ 4443.008 \AA\ and $\lambda$ 4457.431 \AA. Zirconium is solar in HD~1279 (HR~62), HD~123445 (HR~5294), HD~171961 (HR~6990, \textbf{the solar zirconium abundance is -9.42}), 
overabundant in HD~99803A (HR~4423, -8.12), HD~147550 (HR~6096, -8.64), HD~202671 (HR~8137, -7.42), $\mu$ Lep (HR~1702, -7.64) and $\beta$ Scl (HR~8937, -7.32).

\subsection{Barium}

The two resonance lines of \ion{Ba}{2} (Multiplet 1) at 4554.029 \AA\ and 4934.076 \AA\ and the low excitation line 6141.59 \AA\ are present. The hyperfine structure of the various isotopes of barium we have used is collected in Table \ref{tab:Baii}. Barium is solar in HD~1279 (HR~62), HD~123445 (HR~5294), HD~202671 (HR~8137), $\beta$ Scl (HR~8937) and HD~171961 (HR~6990, \textbf{the solar barium abundance is -9.82})
and overabundant in HD~99803A (HR~4423, -8.30), HD~147550 (HR~6096, -8.74) and $\mu$ Lep (HR~1702, -8.82).

\begin{table}
\centering
\caption{The \ion{Ba}{2} linelists}
\begin{tabular}{ccccc}
\hline
Wavelength & 	ion 	   & $\log gf$ & lower energy level & EW \\ 
(\AA) & & & (cm$^{-1}$) & (m\AA) \\ \hline
   4553.934 &  \ion{Zr}{2} &  -0.57 &  19514.840 & 0.1  \\
   4553.995 &  \ion{Ba}{2} &  -1.57 &      0.000 & 0.1 \\
  4553.997  & \ion{Ba}{2}  &  -1.57 &      0.000 &  0.1 \\
  4553.998  & \ion{Ba}{2}  & -1.99  &     0.000  & 0.0  \\
  4553.999  & \ion{Ba}{2}  & -1.82  &     0.000  & 0.1  \\
  4554.001  & \ion{Ba}{2}  & -1.82  &     0.000  &  0.1 \\
  4554.001  & \ion{Ba}{2}  &  -2.22 &      0.000 & 0.0  \\
  4554.011  & \ion{Cr}{1}  &  -0.73 &  33040.093 & 0.0   \\
  4554.029  & \ion{Ba}{2}  &  0.02  &     0.000  & 16.6  \\
  4554.029  & \ion{Ba}{2}  & -1.45  &     0.000  & 1.3  \\
  4554.029  & \ion{Ba}{2}  & -0.94  &     0.000  & 3.8   \\
  4554.046  & \ion{Ba}{2}  & -1.38  &     0.000  & 0.3   \\
  4554.049  & \ion{Ba}{2}  & -1.82  &     0.000  & 0.1  \\
  4554.049  & \ion{Ba}{2}  & -1.15  &     0.000  & 0.6  \\
  4554.050  & \ion{Ba}{2}  & -2.52  &     0.000  & 0.0  \\
  4554.051  & \ion{Ba}{2}  & -1.57  &     0.000  & 0.1  \\
  4554.052  & \ion{Ba}{2}  & -2.29  &     0.000  & 0.0  \\
\hline 
  4934.005  & \ion{Fe}{1}  & -0.61  & 33507.120  &  0.4 \\
  4934.074  & \ion{Ba}{2}  & -3.12  &     0.000  &    0.0 \\
  4934.074  & \ion{Ba}{2}  & -1.33  &     0.000  & 0.8  \\
  4934.075  & \ion{Ba}{2}  & -3.15  &     0.000  & 0.0 \\
  4934.075  & \ion{Ba}{2}  & -1.10  &     0.000  & 1.3  \\
  4934.076  & \ion{Ba}{2}  & -1.77  &     0.000  & 0.3  \\
  4934.076 &  \ion{Ba}{2}  & -1.25  &     0.000  &  0.9  \\
  4934.077 &  \ion{Ba}{2}  & -0.29  &     0.000  &  7.0  \\
  4934.084  & \ion{Fe}{1}  &  -2.30 &  26627.608 &   0.1 \\
\hline
  6141.597  &  \ion{Fe}{1} &   -3.12 &  33765.304 & 0.0  \\
  6141.713  & \ion{Ba}{2}  & -0.22  &  5674.824   &    2.4 \\
  6141.713  & \ion{Ba}{2}  & -1.03  &  5674.824   &     0.4 \\
  6141.714  & \ion{Ba}{2}  & -1.18  &  5674.824   &    0.3   \\
  6141.714  & \ion{Ba}{2}  & -1.26  &  5674.824  & 0.2   \\
  6141.715  & \ion{Ba}{2}  & -1.69  &  5674.824  &  0.1  \\
  6141.717  & \ion{Ba}{2}  & -3.07  &  5674.824  & 0.0  \\
  6141.718  & \ion{Ba}{2}  & -3.05  &  5674.824  & 0.0 \\
  6141.731  & \ion{Fe}{1}  &  -1.61 & 29056.322  & 0.1 \\ \hline
\label{tab:Baii}			      		     	  
\end{tabular}
\end{table} 

\subsection{The Rare Earths}

We have searched for once ionized rare earths elements in the spectra of the eight stars, namely \ion{La}{2}, \ion{Ce}{2}, \ion{Pr}{2}, \ion{Nd}{2}, \ion{Sm}{2}, \ion{Eu}{2}, \ion{Gd}{2}, \ion{Tb}{2}, \ion{Dy}{2}, \ion{Ho}{2} and \ion{Er}{2}. As the twice ionised rare earths often are the dominant stages in these atmospheres of late B stars, we also looked for the twice ionized species, \ion{Pr}{3} and \ion{Nd}{3}, using the lines listed either in NIST or in DREAM and the linelists published in \cite{RRKB} and \cite{2007A&A...473..907R} .
The lines of these ions are weak, a few of these lines could be found in most stars only be found in the slowest rotators. HD~147550 (HR~6096) is the star which has the most detections of rare earths.

\subsubsection{Lanthanum}

The spectrum synthesis of the \ion{La}{2} line at 4042.91 \AA\ feature is consistent with one solar abundance upper limit for HD~171961 (HR~6990) and $\beta$ Scl (HR~8937). The line is detected in HD~1279 (HR~62), HD~99803A (HR~4423) and HD~202671 (HR~8137)
and properly reproduced with a solar abundance (\textbf{the solar lanthanum abundance is -10.90}). Lanthanum is overabundant in HD~147550 (HR~6096, -9.90) and very abundant in $\mu$ Lep (HR~1702, -8.00).

\subsubsection{Cerium}

The synthesis of the unblended  \ion{Ce}{2} lines at 4133.80 \AA\ and 4460.21 \AA\ yields a mild overabundance of cerium  in HD~147550 (HR~6096, -9.24) and a one solar upper limit in all the other stars (\textbf{the solar cerium abundance is -10.42}).
\\
\subsubsection{Praseodymium}

The lines of \ion{Pr}{3} at 5284.69 and 5299.99 \AA\ are unblended and provide large overabundances in HD~147550 (HR~6096, -10.38), HD~99803A (HR~4423, -9.98), $\mu$ Lep (HR~1702, -8.98), 
HD~123445 (HR~5294, -8.58). The synthesis of these lines is consistent with a solar abundance upper value for HD~1279 (HR~62), HD~202671 (HR~8137), HD~171961 (HR~6990) and $\beta$ Scl (HR~8937, \textbf{the solar praseodymium abundance is -11.28})

\subsubsection{Neodymium}

For \ion{Nd}{3}, the strongest expected lines in DREAM are the resonance lines $\lambda$ 5265.019 \AA\ and $\lambda$ 5294.099 \AA. The synthesis of these lines yields moderate to large overabundances of \ion{Nd}{3} in HD~99803A (HR~4423, -9.98), HD~1279 (HR~62, -9.28) and HD~123445 (HR~5294, -8.18), HD~147550 (HR~6096, -9.73). In HD~202671 (HR~8137), HD~171961 (HR~6990), $\mu$ Lep (HR~1702) and $\beta$ Scl (HR~8937) the abundance of neodymium has a \textbf{one solar upper value, ie. -10.50} (the lines are not detected).

\subsubsection{Samarium}

The \ion{Sm}{2} lines at 4280.79 \AA\ and 4424.32 \AA\ are unblended  and are consistent with a 1 solar upper limit abundance of samarium (\textbf{the solar samarium abundance is -11.04}) in all stars, except for HD~147550 (HR~6096) for which samarium is overabundant (-9.44).

\subsubsection{Europium}

The \ion{Eu}{2} resonance line at 4129.73 \AA\ is unblended and consistent with a 1 solar upper limit abundance of europium (\textbf{the solar europium abundance being -11.48}) in all stars except in HD~147550 (HR~6096) where europium is overabundant (-10.48).

\subsubsection{Gadolinium}

The \ion{Gd}{2} line at 4037.32 \AA\ is unblended and properly reproduced with large overabundances in HD~147550 (HR~6096, -9.63) and $\mu$ Lep (HR~1702, -7.45). In the other stars, gadolinium has a one solar upper limit (\textbf{the solar gadolinium abundance is -10.88}).

\subsubsection{Terbium}

The \ion{Tb}{2} line at 4005.47 \AA\ provides a 1 solar upper limit abundance (\textbf{the solar terbium abundance is -11.70}) for all stars except HD~147550 (HR~6096) where terbium is overabundant (-10.70)

\subsubsection{Dysprosium}

The \ion{Dy}{2} line at 4000.45 \AA\ provides a 1 solar upper linit (\textbf{the solar dysprosium abundance is -10.90})  in all stars except for HD~147550 (HR~6096) where dysprosium is overabundant (-9.30).

\subsubsection{Holmium}

The low excitation line of \ion{Ho}{2} at 4152.62 \AA\ provides a 1 solar upper limit (\textbf{the solar holmium abundance is -11.52}) in all stars except for HD~147550 (HR~6096) where holmium is overabundant (-9.04).
\\

\subsubsection{Erbium}

For \ion{Er}{2}, the line at 4142.91 \AA\ is consistent with a 1 solar upper limit (\textbf{the solar erbium abundance is -11.08}) for all stars except HD 147550 (HR~6096) where erbium is very overabundant (-8.78)

 \subsubsection{Thulium}

The line at 4242.15 \AA\ of \ion{Tm}{2} is consistent with a 1 solar upper limit (\textbf{the solar thulium abundance is -11.90}) in all eight stars.
 
\subsubsection{Ytterbium}
 
 The \ion{Yb}{2} line at 4135.095 \AA\ provides a one solar upper limit for all eight stars (\textbf{the solar ytterbium abundance is -11.16}).
 
\subsection{The very heavy elements Hafnium, Osmium, Platinum, Gold and Mercury} 

\subsubsection{Hafnium}

The \ion{Hf}{2} line at 3918.08 \AA\ yields a 1 solar upper limit (\textbf{the solar hafnium abundance being -11.15}) in all eight stars.

\subsubsection{Osmium}

The \ion{Os}{2} line at 4158.44 \AA\ is consistent with a 1 solar upper limit (\textbf{the solar osmium abundance being -10.60}) in all eight stars.

\subsubsection{Platinum}
 
The 4514.47 \AA\ \ion{Pt}{2} line is the only clearly detected line of platinum in the spectra of HD~99803A (HR~4423), $\mu$ Lep (HR~1702) and $\beta$ Scl (HR~8937). In the other five stars, this line is not detected and the abundance of platinum has a one solar upper limit (\textbf{the solar platinum abundance is -10.38}).  
\textbf{Our referee, Prof. M. Dworetsky has kindly provided the hyperfine structure and isotopic shifts for the 4514.47 \AA\ \ion{Pt}{2} line \citep{1989ApJ...340.1140E,1998ApJ...504..533B,1998CoSka..27..324J}. Inclusion of these hyperfine transitions significantly reduces the overabundances of platinum. The new abundances are -8.32 in HR~4423, -7.20 in HR~1702, -6.20 in HR~8937.}
\textbf{There are no other species at this wavelength which could account for the observed absorption}. The wavelength scale around the \ion{Pt}{2} line was checked using the \ion{Fe}{2} neighbouring control lines whose NIST wavelengths are accurate to within $\pm$ 0.001 \AA.



\subsubsection{Gold}

\textbf{The \ion{Au}{2} line at 4052.790 \AA\ is not clearly seen in the spectra of the eight stars and is therefore reproduced by a one solar upper limit for all stars \textbf{(-10.99)}}.
 

\subsubsection{Mercury}

The abundance of mercury has been derived from the low excitation \ion{Hg}{2} line at 3983.93 \AA\ only which is clearly present in all eight stars. This feature absorbs from 2\% in HD~147550 (HR~6096) and HD~202671 (HR~8137) to about 12 \% of the continuum in $\beta$ Scl (HR~8937). The other \ion{Hg}{2} lines are all high excitation lines either weak or blended with more abundant species and were not synthesized. In particular \ion{Hg}{2} 6149.4749 \AA\ is blended with \ion{Fe}{2} 6149.258 \AA\ in all stars. 
\\
To model the \ion{Hg}{2} line at 3983.93 \AA, we have included 9 transitions ie., all hyperfine structure from the various isotopes from $\mathrm{Hg}^{196}$  and $\mathrm{Hg}^{204}$ 
listed in \cite{Do03}. These transitions are collected together with blending lines from \ion{Ti}{1}, \ion{Fe}{1}, \ion{Cr}{1} and \ion{Cr}{2} in Table \ref{tab:Hgii}, \textbf{where the equivalent widths are computed for the final abundances of HR~62}.

\begin{table}
\centering
\caption{The \ion{Hg}{2} 3983.93 \AA\ linelist}
\begin{tabular}{ccccc}
\hline
Wavelength & 	ion 	   & $\log gf$ & lower energy level & EW \\
(\AA) & & & (cm$^{-1}$) & (m\AA) \\ \hline
  3983.771 &  \ion{Hg}{2}  &  -4.50   &    35514.999 &   1.2 \\ 
  3983.827 &   \ion{Ti}{1} & 	-1.46 &   17075.258  &    0.0 \\ 
  3983.832 &   \ion{Fe}{1} & 	-4.81 &   24338.766  &    0.0  \\ 
  3983.839 &   \ion{Hg}{2} &   -3.00  &    35514.000 &   14.1 \\ 
  3983.844 &   \ion{Hg}{2} &   -3.13  &    35514.000 &   12.5 \\ 
  3983.851 &   \ion{Cr}{2} &   -4.31  &    81962.291 &   0.0 \\ 
  3983.853 &   \ion{Hg}{2} &   -3.00  &    35514.000 &   14.1 \\ 
  3983.874 &   \ion{Fe}{1} & 	-2.70 &   34039.515  &  0.0  \\ 
  3983.899 &   \ion{Cr}{1} & 	 0.35 &   20520.904  &  3.2 \\ 
  3983.912 &   \ion{Hg}{2} &   -2.50  &  35514.000   &  19.2 \\ 
  3983.932 &   \ion{Hg}{2} &   -3.10  &  35514.000   & 12.9  \\ 
  3983.941 &   \ion{Hg}{2} &   -2.90  &  35514.000   & 15.3  \\ 
  3983.956 &   \ion{Fe}{1} & 	-1.02 &   21999.129  &  2.5  \\ 
  3983.986 &   \ion{Cr}{2} &   -2.32  &  90512.561   &   0.0  \\ 
  3983.991 &   \ion{Cr}{2} &   -2.88  &  82362.188   &    0.0   \\ 
  3983.993 &   \ion{Hg}{2} &   -3.00  &  35514.000   & 14.2  \\ 
  3984.022 &   \ion{Cr}{1} & 	-2.47 &   35397.971  &  0.0  \\ 
  3984.072 &   \ion{Hg}{2} &   -3.00  &  35514.000   & 14.4  \\ \hline
\label{tab:Hgii}			      		     	  
\end{tabular}
\end{table}

The wavelength scale was checked by using four control lines on each side of the \ion{Hg}{2} line: shortwards the \ion{Zr}{2} line at 3982.0250 \AA\ , the \ion{Y}{2} line at 3982.59 \AA\ and longwards the \ion{Zr}{2} lines at 3984.718 \AA\ and 3991.15 \AA. Once corrected for the radial velocities of each star, the centers of these lines are found at their expected laboratory locations to within $\pm$ 0.02 \AA\  which we will adopt as the accuracy of the wavelength scale in this spectral region.
\\
First, we checked the influence of possible contaminant species to the \ion{Hg}{2} line from 3983.50 \AA\ up to 3984.50 \AA\ which we estimate to be the maximum extension of the line wings. The only possible contaminants are three weak lines: \ion{Fe}{2}  $\lambda$  3983.737 \AA, \ion{Cr}{1} $\lambda$ 3983.897 \AA\  and \ion{Fe}{1} $\lambda$ 3983.956 \AA. The equivalent widths of these lines were computed for the derived Fe and Cr abundances and the sum of their contributions, \textbf{is by far insufficient to reproduce the equivalent width of the observed feature at 3983.93 \AA\ (about 64 to 70  m\AA) of the \ion{Hg}{2} blend}. 
Another test consisted in computing the flux without the \ion{Hg}{2} transitions. This consistently resulted into a very weak absorption feature from 3983.50 \AA\ to 3984.50 \AA\ in each star. We can therefore conclude that the observed features at 3983.93 \AA\ are indeed mostly due to the \ion{Hg}{2} line and almost free of blends.
\\
Including the hyperfine structure of the seven isotopes from Table 2 of \cite{Do03} significantly usually reduces the mercury abundance.
The final combination of oscillator strengths is recorded in Table \ref{tab:Hgii}.
\\
The final mercury abundances derived from the synthesis of the 3983.93 \AA\ line
yield overabundances of -7.43 for HD~147550 (HR~6096) and HD~202671 (HR~8137), of about -7.43 for HD~1219 (HR~62) and HD~99803A (HR~4423), -6.23 in HD~171961 (HR~6990) and of -5.40 for HD~123445 (HR~5294), of -5.23 for $\mu$ Lep (HR~1702) and -4.83 for $\beta$ Scl (HR~8937). \textbf{The solar mercury abundance is -10.91}.

The final abundances \footnote{We here refer to the absolute abundance in the star:
$\log_{10}\left({\mathrm{X}\over\mathrm{H}}\right)_{\star}$ where $\log_{10}\left(\mathrm{H} \right)=12.0$}
are collected in Table \ref{tab:abund_all}. The error budget for these abundances is explained in the Appendix and uncertainties are collected in Table \ref{tab:uncertainties}.

\begin{table}[]
    \centering
    \begin{tabular}{|c|c|c|c|c|c|c|c|c|c|}
         \hline
         Element  & HD\,1279 & HD\,99803A & HD\,123445 & HD\,147550 & HD\,171961 & HD\,202671 & $\mu$\,Lep & $\beta$\,Scl &   Sun   \\ \hline
	          & HR~62    & HR~4423    & HR~5294    & HR~6096    & HR~6990    &  HR~8137   &  HR~1702   &  HR~8937     &          \\ \hline
         He  ( 2) &  10.33   &   10.63    & 10.23      &    10.53   &    9.53    &      9.93  &     9.93   &     9.93     &  10.93  \\
         C   ( 6) &   7.97   &    8.03    &  8.21      &     8.03   &    7.13    &      7.91  &     8.21   &     8.21     &   8.43  \\
         N   ( 7) &   7.83   &            &            &     7.83   &            &            &            &              &   7.83  \\
         O   ( 8) &   8.87   &    8.41    &  8.39      &     8.47   &    8.39    &      8.47  &     8.73   &     8.59     &   8.69  \\
         Ne  (10) &   7.93   &            &            &     7.93   &            &      8.41  &     8.23   &     7.23     &   7.93  \\
         Na  (11) &   6.24   &    5.94    &            &     6.24   &    6.24    &      6.24  &     6.94   &              &   6.24  \\
         Mg  (12) &   7.08   &    6.98    &  7.25      &     7.45   &    6.90    &      7.25  &     7.48   &     7.30     &   7.60  \\
         Al  (13) &   5.45   &    5.93    &  5.85      &     6.75   &    5.45    &      5.15  &     5.45   &     5.15     &   6.45  \\
         Si  (14) &   6.21   &    7.00    &  7.29      &     7.72   &    6.99    &      7.45  &     8.03   &     7.87     &   7.51  \\
         P   (15) &   6.89   &    5.41    &  5.41      &     6.29   &    6.95    &      7.71  &     6.82   &     6.68     &   5.41  \\
         S   (16) &   6.82   &    7.12    &  7.12      &     7.33   &    7.12    &      6.52  &     7.12   &     6.82     &   7.12  \\
         Cl  (17) &   6.50   &            &            &            &            &            &            &              &   5.50  \\
         Ar  (18) &   6.40   &            &            &            &            &      6.40  &            &              &   6.40  \\
         Ca  (20) &   7.04   &    5.64    &  6.12      &     6.22   &    5.34    &      6.64  &     6.74   &     6.74     &   6.34  \\
         Sc  (21) &   3.85   &    3.63    &            &     2.85   &            &            &     3.85   &     2.15     &   3.15  \\
         Ti  (22) &   5.65   &    5.36    &  5.46      &     5.11   &    5.53    &      5.63  &     5.83   &     5.75     &   4.95  \\
         V   (23) &   3.93   &    3.53    &  3.93      &     4.40   &            &            &            &     3.93     &   3.93  \\
         Cr  (24) &   5.88   &    6.27    &  6.04      &     5.78   &    5.64    &      5.94  &     6.36   &     6.29     &   5.64  \\
         Mn  (25) &   6.83   &    6.03    &  7.28      &     5.77   &    5.43    &      6.94  &     7.91   &     7.83     &   5.43  \\
         Fe  (26) &   7.68   &    7.06    &  6.98      &     7.55   &    7.50    &      7.90  &     7.73   &     7.73     &   7.50  \\
         Ni  (28) &   5.52   &    5.52    &  4.92      &     6.48   &    5.52    &      5.46  &     4.76   &     5.52     &   6.22  \\
         Ga  (31) &   4.82   &    3.04    &  6.04      &     3.04   &    5.14    &            &     6.62   &     6.04     &   3.04  \\
         Sr  (38) &   3.47   &    2.97    &  2.87      &     3.87   &    2.87    &      2.87  &     5.57   &     6.05     &   2.87  \\
         Y   (39) &   4.75   &    3.99    &  4.91      &     2.91   &    2.21    &      4.21  &     5.51   &     6.31     &   2.21  \\
         Zr  (40) &   2.58   &    3.88    &  2.58      &     3.36   &    2.58    &      2.58  &     4.36   &     4.68     &   2.58  \\
         Xe  (54) &   5.84   &            &            &     2.24   &            &      4.24  &     6.98   &     6.94     &   2.24  \\
         Ba  (56) &   2.18   &    3.69    &  2.18      &     3.26   &    2.18    &      2.18  &     3.18   &     2.18     &   2.18  \\
         La  (57) &          &    1.10    &            &     2.10   &            &      1.10  &     4.00   &              &   1.10  \\
         Ce  (58) &          &            &            &     2.76   &            &      1.58  &     1.58   &              &   1.58  \\
         Pr  (59) &   0.72   &    2.02    &  3.42      &     1.72   &            &            &     3.02   &              &   0.72  \\
         Nd  (60) &   2.72   &    2.02    &  3.82      &     2.27   &            &            &     1.42   &              &   1.42  \\
         Sm  (62) &          &            &            &     2.56   &            &            &     0.96   &     0.96     &   0.96  \\
         Eu  (63) &          &            &            &     1.52   &            &            &     0.52   &     0.52     &   0.52  \\
         Gd  (64) &          &            &            &     2.37   &            &            &     4.55   &              &   1.07  \\
         Tb  (65) &          &            &            &     1.30   &    0.30    &            &     0.30   &              &   0.30  \\
         Dy  (66) &          &    1.10    &            &     2.70   &    1.10    &            &     1.10   &              &   1.10  \\
         Ho  (67) &          &            &            &     2.96   &            &            &            &     0.48     &   0.48  \\
         Er  (68) &          &            &            &     3.22   &            &            &            &              &   0.92  \\
         Hf  (72) &          &            &            &            &            &            &     0.85   &              &   0.85  \\
         Os  (76) &          &            &            &            &    1.40    &            &            &              &   1.40  \\
         Pt  (78) &          &    1.90    &            &            &    0.02    &            &     2.72   &     4.02     &   0.02  \\
         Hg  (80) &   5.35   &    5.17    &  6.60      &     4.57   &    5.77    &      4.47  &     6.77   &     7.17     &   1.17  \\
         Bi  (83) &   4.95   &            &            &            &            &            &            &              &   0.65  \\  \hline
    \end{tabular}
    \caption{The final abundances for the eight stars}
    \label{tab:abund_all}
\end{table}


\begin{table}
\centering
\caption{Abundance uncertainties for the elements analysed in the eight stars}
\begin{tabular}{ccccccccccc}


\hline
  && HeI & CII&OI&MgII&AlII&SiII &PII &SII& CaII \\ \hline
$\Delta$ \teff  & +200 K & -0.30  & 0.00  & 0.05    & 0.047 &0.075  & -0.017 & 0.00 &  0.00 &  0.109\\
$\Delta \log g$ &  0.15 dex& 0.079  & 0.24  & -0.06   & -0.052   & 0.118  & -0.017 & -0.12  & -0.12 & -0.032\\
$\Delta \xi_{t}$ & +0.20 \kms & 0.00 & 0.00  & 0.00   & 0.00 &  0.075  & -0.035  & -0.05 & -0.046 & 0.00 \\
$\Delta \log gf $ & +0.10     & -0.30 & -0.12  & -0.16  & -0.25 & -0.028   & -0.053 & -0.07 &  -0.071 & -0.105  \\
$\Delta continuum$ &          & -0.12 & 0.21  & 0.075    & 0.13 & 0.25  & 0.062 & 0.06  & 0.061 & 0.058  \\
\hline      
$2\sigma_{[X/H]}$  & & 0.45  & 0.34 &  0.19 & 0.29  & 0.20 & 0.092 & 0.160 & 0.159 & 0.165 \\ 
  \hline
&  &ScII&TiII&VII&CrII&MnII&FeII&NiII&SrII & YII\\ \hline
$\Delta$ \teff  & +200 K & 0.22& 0.041   & 0.028  &  0.057 & 0.00   &  0.079 &  0.176 & 0.138 &0.146 \\
$\Delta \log g$ &  0.15 dex& -0.079& 0.015    & -0.014 &0.03 & 0.007   & 0.041   & 0.114 & -0.058 &  \\
$\Delta \xi_{t}$ & +0.20 \kms & -0.038  &  -0.016 & 0.00 &  0.00 & 0.00   & -0.095& 0.00 & -0.058 & 0.00  \\
$\Delta \log gf $ & +0.10     & -0.13  & -0.23  & -0.046 & -0.15 & -0.015 & -0.095  & -0.155 & -0.058 & -0.05 \\
$\Delta continuum$ &          & 0.067 & 0.03  & 0.014 & 0.046  & 0.014  & 0.0126 & 0.079  & 0.051 &0.041  \\
\hline 
$2\sigma_{[X/H]}$  & & 0.276 & 0.064 & 0.23 & 0.169 & 0.076 & 0.162 & 0.273 & 0.178 &0.165 \\ 
  \hline
&&ZrII&BaII&LaII&CeII&PrIII&NdIII&SmII & EuII&GdII\\ \hline  
$\Delta$ \teff  & +200 K     & 0.058   & 0.114 &  0.11& 0.11 & 0.11 & 0.11 & 0.11 & 0.11 & 0.11 \\
$\Delta \log g$ &  0.15 dex&  0.046    & -0.032   &  -0.125 & -0.13 & -0.13 & -0.13 & -0.13 &  -0.13 & -0.13\\
$\Delta \xi_{t}$ & +0.20 \kms    & 0.00   & -0.097 & -0.09 & -0.09 & -0.09 & -0.09 & -0.09 &-0.09 & -0.09  \\
$\Delta \log gf $ & +0.10         & -0.066  & -0.125 & -0.13   &-0.13  & -0.13  & -0.13 & -0.13 & -0.13 & -0.13\\
$\Delta continuum$ &             & 0.146  & -0.097 & -0.09  & -0.09  & -0.09   & -0.09 & -0.09 &-0.09  & -0.09 \\ 
\hline
$2\sigma_{[X/H]}$  &   & 0.173 & 0.251 & 0.25 & 0.25 & 0.25 & 0.25 & 0.25 &  0.25 & 0.25\\ 
  \hline
&& DyII& TbII&HoII& ErII& Tm II &YbII & HfII & OsII&PtII\\ \hline
$\Delta$ \teff  & +200 K &    0.11 & 0.11 & 0.11 & 0.11 & -0.155 &  0.11 & 0.11 & 0.11& 0.11 \\
$\Delta \log g$ &  0.15 dex & -0.13 & -0.13 & -0.13 & -0.13 & 0.041 & -0.13 & -0.13 & -0.13& -0.13\\
$\Delta \xi_{t}$ & +0.20 \kms & -0.09 & -0.09 & -0.09 & -0.09 & -0.097 & -0.09 & -0.09 & -0.09  & -0.09 \\
$\Delta \log gf $ & +0.10       & -0.13 & -0.13 & -0.13 & -0.13  & -0.108 & -0.13 & -0.13  & -0.13 & -0.13\\
$\Delta continuum$ &            & -0.09 & -0.09 & -0.09 & -0.09  &  0.079 &  -0.09  & -0.09  & -0.09& -0.09\\ 
\hline
$2\sigma_{[X/H]}$  &  & 0.25 & 0.25 & 0.25 & 0.25 & 0.23 & 0.25 & 0.25 & 0.25& 0.25\\ \hline
&&  AuII& \ion{Hg}{2}&   & \\ \hline
$\Delta$ \teff  & +200 K  & 0.11  & -0.155 & &\\
$\Delta \log g$ &  0.15 dex  & -0.13 & 0.041 & &\\
$\Delta \xi_{t}$ & +0.20 \kms & -0.09 & -0.097 &  &\\
$\Delta \log gf $ & +0.10      & -0.13 & -0.108 & &\\
$\Delta continuum$ &            & -0.09 & 0.079 & &\\ 
\hline
$2\sigma_{[X/H]}$  &   & 0.25 & 0.23 & &\\ \hline
\label{tab:uncertainties}
\end{tabular}
\end{table}

\section{Discussion}

\begin{figure}
\epsscale{.80}
\plotone{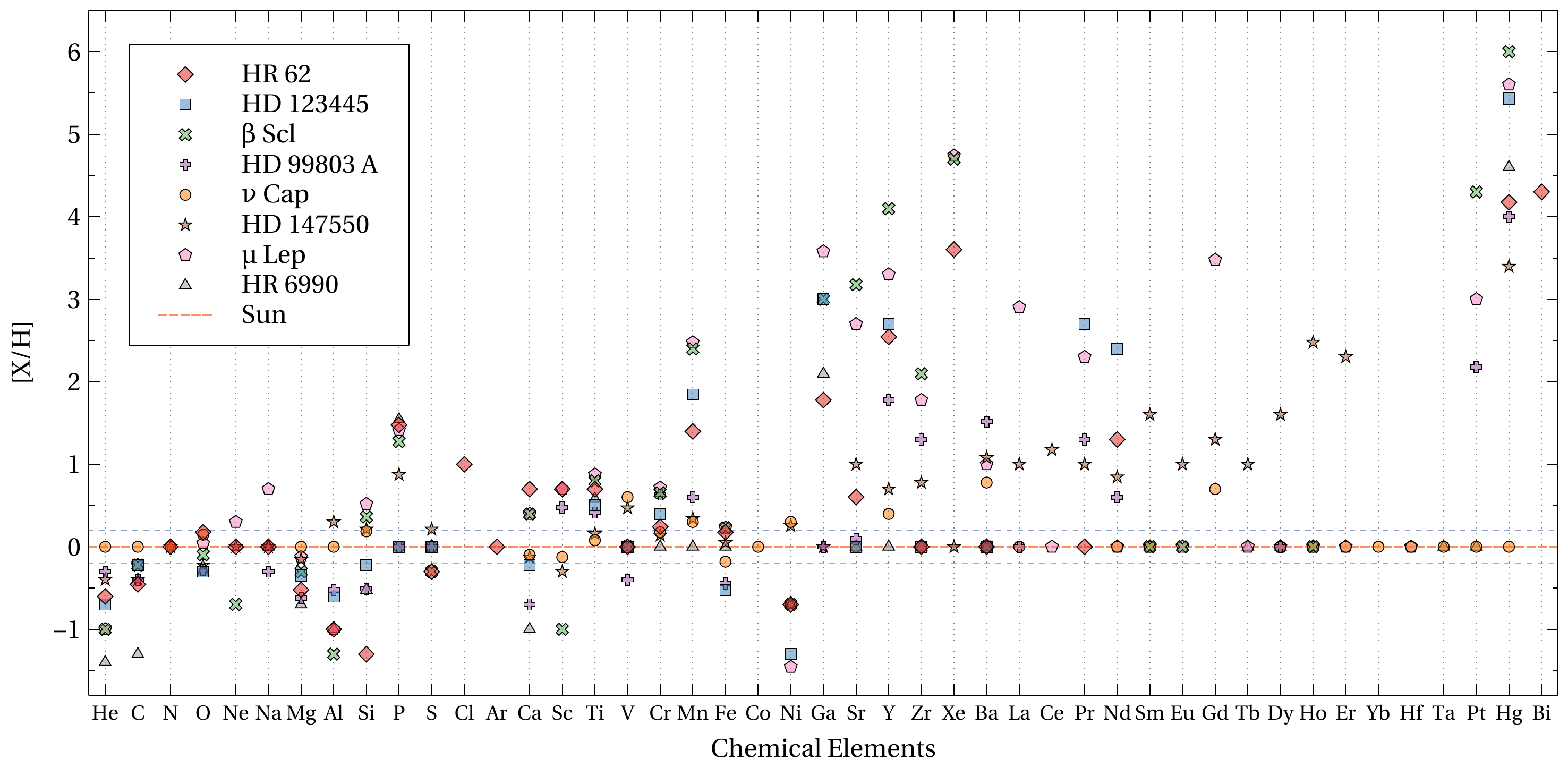}
\caption{Comparison of the abundance patterns of the eight stars to that of the superficially normal star $\nu$ Cap (HR~7773)
\label{abund_patterns}}
\end{figure}

The abundance analysis carried out here clearly establishes that the six slowly rotating southern stars analysed are newly discovered Chemically Peculiar stars with large mercury overabundances. 
The abundance patterns for these objects are shown in Figure \,\ref{abund_patterns}.
None of these objects appears to be a rapid rotator seen pole-on as Vega is. We fail to find the characteristic flat-bottomed \ion{Fe}{2} lines of Vega in the spectra we analysed.
The inference of the age and mass of the stars with the SPInS tool and BaSTI stellar models shows that six stars are still on the main sequence while for the two most massive ones, HD~1279 (HR~62) and HD~202671 (HR~8137), the evolutionary status  cannot be firmly assessed. All these stars are young as expected, their estimated ages range from 100 to 261 Myrs. 
The analysis of the TESS lightcurve of HD~99803A (HR~4423) shows that this star is an eclipsing binary star with an orbital period longer than the 24.96-d time span of the Sector 10 data.
We estimate the radius of the transiting body to be about 0.2 the radius of the primary B9V star from the transit depth. That suggests a secondary that is a lower main-sequence star, a plausible conjecture. There is no evidence for p-modes nor g-modes in the periodogram of the TESS lightcurve. HD~99803 (HR~4423) is too cool to be a SPB star nor a $\beta$ Cephei star. There is no evidence either in this star for signatures of p-modes nor g-modes. HD~123445 (HR~5294) is a pulsational variable with a principal period of 6 hours and 28 minutes and 2 harmonics. 
HD~99803A (HR~4423) stands out as a particularly interesting object also because its spectrum has very sharp lines, similar to HD~72660 (HR~3383). There are very few very sharped lined early A/late B stars.
From the list of identifications provided for the final composition of HD~99803A (HR~4423) in Table 10 in the Appendix, we can conclude that the following species are present in the spectra of HD~99803A (and in the other stars as well): \ion{C}{2}, \ion{N}{2}, \ion{O}{1}, \ion{Na}{1}, \ion{Mg}{2}, \ion{Mg}{1},\ion{Mg}{2}, \ion{Si}{2}, \ion{P}{2}, \ion{S}{2}, \ion{Ca}{2}, \ion{Sc}{2}, \ion{Ti}{2}, \ion{V}{2}, \ion{Cr}{2}, \ion{Cr}{1}, \ion{Mn}{2}, \ion{Fe}{2}, \ion{Ni}{2}, \ion{Ga}{2}, \ion{Sr}{2}, \ion{Y}{2}, \ion{Zr}{2}, \ion{Ba}{2}, \ion{Pt}{2}, \ion{Hg}{2}. Among these species, \ion{Si}{2}, \ion{Ti}{2}, \ion{Cr}{2}, \ion{Mn}{2}, \ion{Fe}{2}, \ion{Y}{2}, \ion{Zr}{2}, have numerous and strong lines and are important opacity sources. Lines from neutrals: \ion{Cr}{1}, \ion{Mn}{1}, \ion{Fe}{1} are also observed and yield lower abundances than \ion{Cr}{2}, \ion{Mn}{2} and \ion{Fe}{2}, indicating possibly departures from LTE.
\\
For the Rare Earths, we find evidence for overbundances of four Rare Earths from unblended lines of \ion{Ce}{2}, \ion{Pr}{3} and \ion{Nd}{3} and \ion{Gd}{2} in the slowest rotating stars. For the other Rare Earths, we only find upper limits: the abundances must be solar or lower.  
The presence of once ionised Rare Earths is difficult to assess at these temperatures above 10000 K
because they are not the dominant ionisation stage. In general, the abundances for the Rare Earths should be regarded as the least reliable ones in this study because they were usually infered from the synthesis of one weak line of the once ionised species only.
\\
The overabundances found in the six newly discovered CP stars run from very mild (2.0 $\odot$  for Fe) up to quite large 
($10^{5}$ $\odot$ for Hg). The underabundances run from mild -0.80 $\odot$ for carbon to pronounced -0.02 $\odot$ for nickel. Helium is quite underabundant in HD 171961 (HR~6990), for which the abundance is only 4 \% of the solar abundance. 
The comparison late B star, $\nu$ Cap (HR~7773, HD~193432), has a nearly solar composition for all elements as already found by  \cite{2018ApJ...854...50M} and \cite{1991MNRAS.252..116A}. For the other eight stars, the overall trend is that the light elements (He through Mg) are most often deficient whereas elements heavier than manganese tend to be more and more overabundant as the atomic mass increases (with the exception of nickel).
This pattern of abundances is most likely caused by radiative diffusion.

\section{Conclusion}

With their characteristic underabundances of most light elements up to calcium and overabundances of manganese and elements heavier than strontium, in particular mercury, the six stars studied here are definitely new HgMn stars or new PGa stars. Their abundances differ from those of the classical HgMn stars $\mu$ Lep (HR~1702) and $\beta$ Scl (HR~8937) possibly because their initial abundances differed slightly and because radiative diffusion operated in a slightly different manner in stars having different masses. For HD~147550 (HR~6096), the mild deficiencies for light elements and overabundances for strontium, yttrium and zirconium and several rare-earths and mercury suggest that this object should be reclassified as a mild Chemically Peculiar star. We find that HD~123445 (HR~5294) pulsates with a main period of 6 hours 28 minutes and two harmonics. More observations of this interesting star will help elucidate its nature. A photometric monitoring of HD~99803A (HR~4423) over a period of time longer than 25 days span of the TESS observations should help measure the orbital period of this new eclipsing spectroscopic binary. 

\label{sec:conclusion}
\begin{acknowledgements} 
\textbf{We thank Prof. Mike Dworetsky, our referee, for providing the hyperfine structure and isotopic shifts for the \ion{Pt}{2} line at 4514 \AA.} We also thank Prof. Charles Cowley for his insightful comments during the analysis of HD~99803A.
We are grateful to Tamas Borkovits, Tibor Mitnyan, and Veselin Kostov for their help with some of the {\it TESS} data.
This work has made use of the VALD database, operated at Uppsala University, the Institute of Astronomy RAS in Moscow, and the University of Vienna.
We have also used the NIST Atomic Spectra Database (version 5.4) available at \url{http://physics.nist.gov/asd}. This research has made use of the SIMBAD database, operated at CDS, 
Strasbourg, France. DMB gratefully acknowledges funding from the Research Foundation Flanders (FWO) by means of a senior postdoctoral fellowship (grant agreement No. 1286521N).
This work was supported by FCT/MCTES through the research grants 
UIDB/04434/2020, UIDP/04434/2020 and PTDC/FIS-AST/30389/2017. MD is 
supported by national funds through FCT in the form of a work contract.
\textbf{Some of the observations reported in this paper were obtained with the 
Southern African Large Telescope (SALT) under programs: 2019-1-MLT-006, 
2019-2-SCI-011, 2020-1-SCI-014, 2020-2-SCI-007, and 2021-1-MLT-004 (PI: 
Ewa Niemczura). Polish participation in SALT is funded by grant No. MEiN 
nr 2021/WK/01. E.N. is supported by the IDUB project held at the 
University of Wroclaw.}

\end{acknowledgements}



\bibliography{ref.bib}

\begin{thebibliography}{90}
\expandafter\ifx\csname natexlab\endcsname\relax\def\natexlab#1{#1}\fi

\bibitem[{{Adelman}(1991)}]{1991MNRAS.252..116A}
{Adelman}, S.~J. 1991, \mnras, 252, 116

\bibitem[{{Adelman} {et~al.}(1993){Adelman}, {Cowley}, {Leckrone}, {Roby}, \&
  {Wahlgren}}]{1993ApJ...419..276A}
{Adelman}, S.~J., {Cowley}, C.~R., {Leckrone}, D.~S., {Roby}, S.~W., \&
  {Wahlgren}, G.~M. 1993, \apj, 419, 276

\bibitem[{{Adelman} \& {Pintado}(1997)}]{1997A&AS..125..219A}
{Adelman}, S.~J. \& {Pintado}, O.~I. 1997, \aaps, 125, 219

\bibitem[{{Bohlender} {et~al.}(1998){Bohlender}, {Dworetsky}, \&
  {Jomaron}}]{1998ApJ...504..533B}
{Bohlender}, D.~A., {Dworetsky}, M.~M., \& {Jomaron}, C.~M. 1998, \apj, 504,
  533

\bibitem[{{Caffau} {et~al.}(2011){Caffau}, {Ludwig}, {Steffen}, {Freytag}, \&
  {Bonifacio}}]{Caffau2011}
{Caffau}, E., {Ludwig}, H.~G., {Steffen}, M., {Freytag}, B., \& {Bonifacio}, P.
  2011, \solphys, 268, 255

\bibitem[{{Carlsson}(1986)}]{Carlsson86}
{Carlsson}, M. 1986, Uppsala Astronomical Observatory Reports, 33

\bibitem[{{Carlsson}(1992)}]{Carlsson92}
{Carlsson}, M. 1992, in Astronomical Society of the Pacific Conference Series,
  Vol.~26, Cool Stars, Stellar Systems, and the Sun, ed. M.~S. {Giampapa} \&
  J.~A. {Bookbinder}, 499

\bibitem[{{Castelli} \& {Hubrig}(2004)}]{Castelli2004}
{Castelli}, F. \& {Hubrig}, S. 2004, \aap, 425, 263

\bibitem[{{Castelli} \& {Kurucz}(2003)}]{Castelli2003}
{Castelli}, F. \& {Kurucz}, R.~L. 2003, in IAU Symposium, Vol. 210, Modelling
  of Stellar Atmospheres, ed. N.~{Piskunov}, W.~W. {Weiss}, \& D.~F. {Gray},
  A20

\bibitem[{{Catanzaro} \& {Leone}(2002)}]{2002NewA....7..495C}
{Catanzaro}, G. \& {Leone}, F. 2002, \na, 7, 495

\bibitem[{{Corbally}(1984)}]{1984ApJS...55..657C}
{Corbally}, C.~J. 1984, \apjs, 55, 657

\bibitem[{{Cowley} {et~al.}(2007){Cowley}, {Hubrig}, {Castelli},
  {Gonz{\'a}lez}, \& {Wolff}}]{2007MNRAS.377.1579C}
{Cowley}, C.~R., {Hubrig}, S., {Castelli}, F., {Gonz{\'a}lez}, J.~F., \&
  {Wolff}, B. 2007, \mnras, 377, 1579

\bibitem[{{Crause} {et~al.}(2014){Crause}, {Sharples}, {Bramall}, {Schmoll},
  {Clark}, {Younger}, {Tyas}, {Ryan}, {Brink}, {Strydom}, {Buckley},
  {Wilkinson}, {Crawford}, \& {Depagne}}]{SALT14}
{Crause}, L.~A., {Sharples}, R.~M., {Bramall}, D.~G., {et~al.} 2014, in Society
  of Photo-Optical Instrumentation Engineers (SPIE) Conference Series, Vol.
  9147, Ground-based and Airborne Instrumentation for Astronomy V, ed. S.~K.
  {Ramsay}, I.~S. {McLean}, \& H.~{Takami}, 91476T

\bibitem[{{Cunto} \& {Mendoza}(1992)}]{1992RMxAA..23..107C}
{Cunto}, W. \& {Mendoza}, C. 1992, \rmxaa, 23

\bibitem[{{Deal} {et~al.}(2018){Deal}, {Alecian}, {Lebreton}, {Goupil},
  {Marques}, {LeBlanc}, {Morel}, \& {Pichon}}]{deal18}
{Deal}, M., {Alecian}, G., {Lebreton}, Y., {et~al.} 2018, \aap, 618, A10

\bibitem[{{Den Hartog} {et~al.}(2003){Den Hartog}, {Lawler}, {Sneden}, \&
  {Cowan}}]{2003ApJS..148..543D}
{Den Hartog}, E.~A., {Lawler}, J.~E., {Sneden}, C., \& {Cowan}, J.~J. 2003,
  \apjs, 148, 543

\bibitem[{{Dimitrijevic} \& {Sahal-Brechot}(1984)}]{Dimitrijevic1984}
{Dimitrijevic}, M.~S. \& {Sahal-Brechot}, S. 1984, \aap, 136, 289

\bibitem[{{Dolk} {et~al.}(2003){Dolk}, {Wahlgren}, \& {Hubrig}}]{Do03}
{Dolk}, L., {Wahlgren}, G.~M., \& {Hubrig}, S. 2003, \aap, 402, 299

\bibitem[{{Drawin}(1969)}]{drawin1969}
{Drawin}, H.~W. 1969, Zeitschrift fur Physik, 225, 483

\bibitem[{{Dworetsky} \& {Budaj}(2000)}]{2000MNRAS.318.1264D}
{Dworetsky}, M.~M. \& {Budaj}, J. 2000, \mnras, 318, 1264

\bibitem[{{Dworetsky} {et~al.}(1998){Dworetsky}, {Jomaron}, \& {Smith}}]{Dwo98}
{Dworetsky}, M.~M., {Jomaron}, C.~M., \& {Smith}, C.~A. 1998, \aap, 333, 665

\bibitem[{{Dworetsky} {et~al.}(2008){Dworetsky}, {Persaud}, \&
  {Patel}}]{2008MNRAS.385.1523D}
{Dworetsky}, M.~M., {Persaud}, J.~L., \& {Patel}, K. 2008, \mnras, 385, 1523

\bibitem[{{Ekstr{\"o}m} {et~al.}(2012){Ekstr{\"o}m}, {Georgy}, {Eggenberger},
  {Meynet}, {Mowlavi}, {Wyttenbach}, {Granada}, {Decressin}, {Hirschi},
  {Frischknecht}, {Charbonnel}, \& {Maeder}}]{Ekstrom2012}
{Ekstr{\"o}m}, S., {Georgy}, C., {Eggenberger}, P., {et~al.} 2012, \aap, 537,
  A146

\bibitem[{{Engleman}(1989)}]{1989ApJ...340.1140E}
{Engleman}, Rolf, J. 1989, \apj, 340, 1140

\bibitem[{{Fuhr} {et~al.}(1988){Fuhr}, {Martin}, \& {Wiese}}]{Fuhr1988}
{Fuhr}, J.~R., {Martin}, G.~A., \& {Wiese}, W.~L. 1988, Journal of Physical and
  Chemical Reference Data, 17

\bibitem[{{Garrison} \& {Gray}(1994)}]{1994AJ....107.1556G}
{Garrison}, R.~F. \& {Gray}, R.~O. 1994, \aj, 107, 1556

\bibitem[{{Grevesse} \& {Sauval}(1998)}]{1998SSRv...85..161G}
{Grevesse}, N. \& {Sauval}, A.~J. 1998, \ssr, 85, 161

\bibitem[{Guitou {et~al.}(2015)Guitou, Spielfiedel, Rodionov, Yakovleva,
  Belyaev, Merle, Thévenin, \& Feautrier}]{GUITOU201594}
Guitou, M., Spielfiedel, A., Rodionov, D., {et~al.} 2015, Chemical Physics,
  462, 94 , inelastic Processes in Atomic, Molecular and Chemical Physics

\bibitem[{{Hauck} \& {Mermilliod}(1998)}]{1998A&AS..129..431H}
{Hauck}, B. \& {Mermilliod}, M. 1998, \aaps, 129, 431

\bibitem[{{Hidalgo} {et~al.}(2018){Hidalgo}, {Pietrinferni}, {Cassisi},
  {Salaris}, {Mucciarelli}, {Savino}, {Aparicio}, {Silva Aguirre}, \&
  {Verma}}]{Hidalgo2018}
{Hidalgo}, S.~L., {Pietrinferni}, A., {Cassisi}, S., {et~al.} 2018, \apj, 856,
  125

\bibitem[{{Holt} {et~al.}(1999){Holt}, {Scholl}, \&
  {Rosner}}]{1999MNRAS.306..107H}
{Holt}, R.~A., {Scholl}, T.~J., \& {Rosner}, S.~D. 1999, \mnras, 306, 107

\bibitem[{{Hubeny} \& {Lanz}(1992)}]{Hubeny}
{Hubeny}, I. \& {Lanz}, T. 1992, \aap, 262, 501

\bibitem[{{Jomaron} {et~al.}(1998){Jomaron}, {Dworetsky}, \&
  {Bohlender}}]{1998CoSka..27..324J}
{Jomaron}, C.~M., {Dworetsky}, M.~M., \& {Bohlender}, D.~A. 1998, Contributions
  of the Astronomical Observatory Skalnate Pleso, 27, 324

\bibitem[{{Kochukhov} {et~al.}(2021){Kochukhov}, {Khalack}, {Kobzar}, {Neiner},
  {Paunzen}, {Labadie-Bartz}, \& {David-Uraz}}]{2021MNRAS.506.5328K}
{Kochukhov}, O., {Khalack}, V., {Kobzar}, O., {et~al.} 2021, \mnras, 506, 5328

\bibitem[{{Kramida} {et~al.}(2018){Kramida}, {Ralchenko}, {Reader}, \& {NIST
  ASD Team}}]{Kramida2018}
{Kramida}, A., {Ralchenko}, Y., {Reader}, J., \& {NIST ASD Team}. 2018,
  http://physics.nist.gov/asd, 5

\bibitem[{{Kurtz}(1985)}]{1985MNRAS.213..773K}
{Kurtz}, D.~W. 1985, \mnras, 213, 773

\bibitem[{{Kurucz}(1992)}]{Kurucz}
{Kurucz}, R.~L. 1992, \rmxaa, 23

\bibitem[{{Kurucz}(2005)}]{atlas12}
{Kurucz}, R.~L. 2005, Memorie della Societa Astronomica Italiana Supplementi,
  8, 14

\bibitem[{{Kurucz}(2013)}]{atlas12_2}
{Kurucz}, R.~L. 2013, {ATLAS12: Opacity sampling model atmosphere program},
  Astrophysics Source Code Library

\bibitem[{{Kurucz} \& {Avrett}(1981)}]{1981SAOSR.391.....K}
{Kurucz}, R.~L. \& {Avrett}, E.~H. 1981, SAO Special Report, 391

\bibitem[{{Lanz} {et~al.}(1993){Lanz}, {Artru}, {Didelon}, \&
  {Mathys}}]{Lanz1993}
{Lanz}, T., {Artru}, M.~C., {Didelon}, P., \& {Mathys}, G. 1993, \aap, 272, 465

\bibitem[{{Lanz} {et~al.}(1988){Lanz}, {Dimitrijevic}, \&
  {Artru}}]{1988A&A...192..249L}
{Lanz}, T., {Dimitrijevic}, M.~S., \& {Artru}, M.-C. 1988, \aap, 192, 249

\bibitem[{{Lawler} {et~al.}(2006){Lawler}, {Den Hartog}, {Sneden}, \&
  {Cowan}}]{LA06}
{Lawler}, J.~E., {Den Hartog}, E.~A., {Sneden}, C., \& {Cowan}, J.~J. 2006,
  \apjs, 162, 227

\bibitem[{{Lawler} {et~al.}(2001{\natexlab{a}}){Lawler}, {Wickliffe}, {Cowley},
  \& {Sneden}}]{law01}
{Lawler}, J.~E., {Wickliffe}, M.~E., {Cowley}, C.~R., \& {Sneden}, C.
  2001{\natexlab{a}}, \apjs, 137, 341

\bibitem[{{Lawler} {et~al.}(2001{\natexlab{b}}){Lawler}, {Wickliffe}, {den
  Hartog}, \& {Sneden}}]{LWHS}
{Lawler}, J.~E., {Wickliffe}, M.~E., {den Hartog}, E.~A., \& {Sneden}, C.
  2001{\natexlab{b}}, Astrophys. J., 563, 1075, (LWHS)

\bibitem[{{Lebreton} \& {Reese}(2020)}]{Lebreton2020L}
{Lebreton}, Y. \& {Reese}, D.~R. 2020, \aap, 642, A88

\bibitem[{{Leone} \& {Manfre}(1997)}]{1997A&A...320..257L}
{Leone}, F. \& {Manfre}, M. 1997, \aap, 320, 257

\bibitem[{{Lodders}(2010)}]{Lodders2010}
{Lodders}, K. 2010, in Astrophysics and Space Science Proceedings, Vol.~16,
  Principles and Perspectives in Cosmochemistry, 379

\bibitem[{{Marques} {et~al.}(2013){Marques}, {Goupil}, {Lebreton}, {Talon},
  {Palacios}, {Belkacem}, {Ouazzani}, {Mosser}, {Moya}, {Morel}, {Pichon},
  {Mathis}, {Zahn}, {Turck-Chi{\`e}ze}, \& {Nghiem}}]{marques13}
{Marques}, J.~P., {Goupil}, M.~J., {Lebreton}, Y., {et~al.} 2013, \aap, 549,
  A74

\bibitem[{{Martin} {et~al.}(1988){Martin}, {Fuhr}, \& {Wiese}}]{Martin88}
{Martin}, G.~A., {Fuhr}, J.~R., \& {Wiese}, W.~L. 1988, {Atomic transition
  probabilities. Scandium through Manganese}

\bibitem[{{Merle} {et~al.}(2011){Merle}, {Th{\'e}venin}, {Pichon}, \&
  {Bigot}}]{2011MNRAS.418..863M}
{Merle}, T., {Th{\'e}venin}, F., {Pichon}, B., \& {Bigot}, L. 2011, \mnras,
  418, 863

\bibitem[{{Monier} {et~al.}(2015){Monier}, {Gebran}, \& {Royer}}]{Monier}
{Monier}, R., {Gebran}, M., \& {Royer}, F. 2015, \aap, 577, A96

\bibitem[{{Monier} {et~al.}(2018){Monier}, {Gebran}, {Royer}, {Kilicoglu}, \&
  {Fr{\'e}mat}}]{2018ApJ...854...50M}
{Monier}, R., {Gebran}, M., {Royer}, F., {Kilicoglu}, T., \& {Fr{\'e}mat}, Y.
  2018, \apj, 854, 50

\bibitem[{{Monier} {et~al.}(2019){Monier}, {Griffin}, {Gebran},
  {K{\i}l{\i}{\c{c}}o{\u{g}}lu}, {Merle}, \& {Royer}}]{2019AJ....158..157M}
{Monier}, R., {Griffin}, E., {Gebran}, M., {et~al.} 2019, \aj, 158, 157

\bibitem[{{Monier} {et~al.}(2020){Monier}, {Niemczura}, \&
  {K{\i}l{\i}{\c{c}}o{\u{g}}lu}}]{2020RNAAS...4..234M}
{Monier}, R., {Niemczura}, E., \& {K{\i}l{\i}{\c{c}}o{\u{g}}lu}, T. 2020,
  Research Notes of the American Astronomical Society, 4, 234

\bibitem[{{Moon} \& {Dworetsky}(1985)}]{1985MNRAS.217..305M}
{Moon}, T.~T. \& {Dworetsky}, M.~M. 1985, \mnras, 217, 305

\bibitem[{{Morel} \& {Lebreton}(2008)}]{morel08}
{Morel}, P. \& {Lebreton}, Y. 2008, \apss, 316, 61

\bibitem[{{Napiwotzki} {et~al.}(1993){Napiwotzki}, {Schoenberner}, \&
  {Wenske}}]{Napiwotzki}
{Napiwotzki}, R., {Schoenberner}, D., \& {Wenske}, V. 1993, \aap, 268, 653

\bibitem[{{Nielsen} {et~al.}(2000){Nielsen}, {Karlsson}, \&
  {Wahlgren}}]{2000A&A...363..815N}
{Nielsen}, K., {Karlsson}, H., \& {Wahlgren}, G.~M. 2000, \aap, 363, 815

\bibitem[{{Nielsen} {et~al.}(2005){Nielsen}, {Wahlgren}, {Proffitt},
  {Leckrone}, \& {Adelman}}]{2005AJ....130.2312N}
{Nielsen}, K.~E., {Wahlgren}, G.~M., {Proffitt}, C.~R., {Leckrone}, D.~S., \&
  {Adelman}, S.~J. 2005, \aj, 130, 2312

\bibitem[{{Nilsson} {et~al.}(2006){Nilsson}, {Ljung}, {Lundberg}, \&
  {Nielsen}}]{2006A&A...445.1165N}
{Nilsson}, H., {Ljung}, G., {Lundberg}, H., \& {Nielsen}, K.~E. 2006, \aap,
  445, 1165

\bibitem[{{P{\'a}l}(2012)}]{Pal12}
{P{\'a}l}, A. 2012, \mnras, 421, 1825

\bibitem[{{Perruchot} {et~al.}(2008){Perruchot}, {Kohler}, {Bouchy}, {Richaud},
  {Richaud}, {Moreaux}, {Merzougui}, {Sottile}, {Hill}, {Knispel}, {Regal},
  {Meunier}, {Ilovaisky}, {Le Coroller}, {Gillet}, {Schmitt}, {Pepe}, {Fleury},
  {Sosnowska}, {Vors}, {M{\'e}gevand}, {Blanc}, {Carol}, {Point}, {Laloge}, \&
  {Brunel}}]{Perruchot}
{Perruchot}, S., {Kohler}, D., {Bouchy}, F., {et~al.} 2008, in \procspie, Vol.
  7014, Ground-based and Airborne Instrumentation for Astronomy II, 70140J

\bibitem[{{Ricker} {et~al.}(2015){Ricker}, {Winn}, {Vanderspek}, {Latham},
  {Bakos}, {Bean}, {Berta-Thompson}, {Brown}, {Buchhave}, {Butler}, {Butler},
  {Chaplin}, {Charbonneau}, {Christensen-Dalsgaard}, {Clampin}, {Deming},
  {Doty}, {De Lee}, {Dressing}, {Dunham}, {Endl}, {Fressin}, {Ge}, {Henning},
  {Holman}, {Howard}, {Ida}, {Jenkins}, {Jernigan}, {Johnson}, {Kaltenegger},
  {Kawai}, {Kjeldsen}, {Laughlin}, {Levine}, {Lin}, {Lissauer}, {MacQueen},
  {Marcy}, {McCullough}, {Morton}, {Narita}, {Paegert}, {Palle}, {Pepe},
  {Pepper}, {Quirrenbach}, {Rinehart}, {Sasselov}, {Sato}, {Seager},
  {Sozzetti}, {Stassun}, {Sullivan}, {Szentgyorgyi}, {Torres}, {Udry}, \&
  {Villasenor}}]{2015JATIS...1a4003R}
{Ricker}, G.~R., {Winn}, J.~N., {Vanderspek}, R., {et~al.} 2015, Journal of
  Astronomical Telescopes, Instruments, and Systems, 1, 014003

\bibitem[{{Roby} \& {Lambert}(1990)}]{1990ApJS...73...67R}
{Roby}, S.~W. \& {Lambert}, D.~L. 1990, \apjs, 73, 67

\bibitem[{{Roby} {et~al.}(1999){Roby}, {Leckrone}, \&
  {Adelman}}]{1999ApJ...524..974R}
{Roby}, S.~W., {Leckrone}, D.~S., \& {Adelman}, S.~J. 1999, \apj, 524, 974

\bibitem[{{Royer} {et~al.}(2014){Royer}, {Gebran}, {Monier}, {Adelman},
  {Smalley}, {Pintado}, {Reiners}, {Hill}, \& {Gulliver}}]{Royer}
{Royer}, F., {Gebran}, M., {Monier}, R., {et~al.} 2014, \aap, 562, A84

\bibitem[{{Ryabchikova} {et~al.}(2015){Ryabchikova}, {Piskunov}, {Kurucz},
  {Stempels}, {Heiter}, {Pakhomov}, \& {Barklem}}]{rya2015}
{Ryabchikova}, T., {Piskunov}, N., {Kurucz}, R.~L., {et~al.} 2015, \physscr,
  90, 054005

\bibitem[{{Ryabchikova} {et~al.}(2006){Ryabchikova}, {Ryabtsev}, {Kochukhov},
  \& {Bagnulo}}]{RRKB}
{Ryabchikova}, T., {Ryabtsev}, A., {Kochukhov}, O., \& {Bagnulo}, S. 2006,
  Astron. and Astrophys., 456, 329, (RRKB)

\bibitem[{{Ryabchikova} {et~al.}(2007){Ryabchikova}, {Sachkov}, {Kochukhov}, \&
  {Lyashko}}]{2007A&A...473..907R}
{Ryabchikova}, T., {Sachkov}, M., {Kochukhov}, O., \& {Lyashko}, D. 2007, \aap,
  473, 907

\bibitem[{{Sakhibullin}(1987)}]{1987SvA....31..151S}
{Sakhibullin}, N.~A. 1987, \sovast, 31, 151

\bibitem[{{Salpeter}(1955)}]{Salpeter1955}
{Salpeter}, E.~E. 1955, \apj, 121, 161

\bibitem[{{Savanov} \& {Hubrig}(2003)}]{2003ASPC..305..230S}
{Savanov}, I. \& {Hubrig}, S. 2003, in Astronomical Society of the Pacific
  Conference Series, Vol. 305, Magnetic Fields in O, B and A Stars: Origin and
  Connection to Pulsation, Rotation and Mass Loss, ed. L.~A. {Balona}, H.~F.
  {Henrichs}, \& R.~{Medupe}, 230

\bibitem[{{Seaton}(1962)}]{seaton62}
{Seaton}, M.~J. 1962, Proceedings of the Physical Society, 79, 1105

\bibitem[{{Shamey}(1969)}]{shamey}
{Shamey}, L.~J. 1969, PhD thesis, UNIVERSITY OF COLORADO AT BOULDER.

\bibitem[{{Smalley} {et~al.}(2017){Smalley}, {Antoci}, {Holdsworth}, {Kurtz},
  {Murphy}, {De Cat}, {Anderson}, {Catanzaro}, {Collier Cameron}, {Hellier},
  {Maxted}, {Norton}, {Pollacco}, {Ripepi}, {West}, \&
  {Wheatley}}]{2017MNRAS.465.2662S}
{Smalley}, B., {Antoci}, V., {Holdsworth}, D.~L., {et~al.} 2017, \mnras, 465,
  2662

\bibitem[{{Smith}(1994)}]{Smith94}
{Smith}, K.~C. 1994, \aap, 291, 521

\bibitem[{{Smith}(1996)}]{Smith96}
{Smith}, K.~C. 1996, \apss, 237, 77

\bibitem[{{Smith}(1997)}]{Smith97}
{Smith}, K.~C. 1997, \aap, 319, 928

\bibitem[{{Smith} \& {Dworetsky}(1993)}]{smith1993}
{Smith}, K.~C. \& {Dworetsky}, M.~M. 1993, \aap, 274, 335

\bibitem[{{Stuik} {et~al.}(2017){Stuik}, {Bailey}, {Dorval}, {Talens},
  {Laginja}, {Mellon}, {Lomberg}, {Crawford}, {Ireland}, {Mamajek}, \&
  {Kenworthy}}]{2017A&A...607A..45S}
{Stuik}, R., {Bailey}, J.~I., {Dorval}, P., {et~al.} 2017, \aap, 607, A45

\bibitem[{{Takeda} {et~al.}(2014){Takeda}, {Kawanomoto}, \&
  {Ohishi}}]{2014PASJ...66...23T}
{Takeda}, Y., {Kawanomoto}, S., \& {Ohishi}, N. 2014, \pasj, 66, 23

\bibitem[{{Takeda} {et~al.}(1999){Takeda}, {Takada-Hidai}, {Jugaku}, {Sakaue},
  \& {Sadakane}}]{1999PASJ...51..961T}
{Takeda}, Y., {Takada-Hidai}, M., {Jugaku}, J., {Sakaue}, A., \& {Sadakane}, K.
  1999, \pasj, 51, 961

\bibitem[{{Talens} {et~al.}(2018){Talens}, {Deul}, {Stuik}, {Burggraaff},
  {Lesage}, {Spronck}, {Mellon}, {Bailey}, {Mamajek}, {Kenworthy}, \&
  {Snellen}}]{2018A&A...619A.154T}
{Talens}, G.~J.~J., {Deul}, E.~R., {Stuik}, R., {et~al.} 2018, \aap, 619, A154

\bibitem[{{Talens} {et~al.}(2017){Talens}, {Spronck}, {Lesage}, {Otten},
  {Stuik}, {Pollacco}, \& {Snellen}}]{2017A&A...601A..11T}
{Talens}, G.~J.~J., {Spronck}, J.~F.~P., {Lesage}, A.~L., {et~al.} 2017, \aap,
  601, A11

\bibitem[{{Thiam} {et~al.}(2010){Thiam}, {LeBlanc}, {Khalack}, \&
  {Wade}}]{2010MNRAS.405.1384T}
{Thiam}, M., {LeBlanc}, F., {Khalack}, V., \& {Wade}, G.~A. 2010, \mnras, 405,
  1384

\bibitem[{{Vidal} {et~al.}(1973){Vidal}, {Cooper}, \&
  {Smith}}]{1973ApJS...25...37V}
{Vidal}, C.~R., {Cooper}, J., \& {Smith}, E.~W. 1973, \apjs, 25, 37

\bibitem[{{Wiese} {et~al.}(1996){Wiese}, {Fuhr}, \& {Deters}}]{W96}
{Wiese}, W.~L., {Fuhr}, J.~R., \& {Deters}, T.~M. 1996, {Atomic transition
  probabilities of carbon, nitrogen, and oxygen : a critical data compilation}

\bibitem[{{Woolf} \& {Lambert}(1999)}]{1999ApJ...521..414W}
{Woolf}, V.~M. \& {Lambert}, D.~L. 1999, \apj, 521, 414

\bibitem[{{Zdravkov} \& {Pamyatnykh}(2008)}]{2008JPhCS.118a2079Z}
{Zdravkov}, T. \& {Pamyatnykh}, A.~A. 2008, in Journal of Physics Conference
  Series, Vol. 118, Journal of Physics Conference Series, 012079

\end{thebibliography}
\bibliographystyle{aa}




\newpage



\begin{appendix}

\section{Determination of uncertainties}
For a representative line of a given element, six major sources are included in the uncertainty determinations: the uncertainty on the effective temperature ($\sigma_{T_{\rm{eff}}}$), on the surface gravity ($\sigma_{\log g}$), on the microturbulent velocity ($\sigma_{\xi_{t}}$), on the apparent rotational velocity ($\sigma_{v_{e}\sin i}$), the oscillator strength ($\sigma_{\log gf}$) and the continuum placement ($\sigma_{cont}$). These uncertainties are supposed to be independent, so that the total uncertainty $\sigma_{tot_{i}}$ for a given transition (i) verifies:\\
\begin{equation}
\sigma_{tot_{i}}^{2}=\sigma_{T_{\rm{eff}}}^{2}+\sigma_{\log g}^{2}+\sigma_{\xi_{t}}^{2}+\sigma_{v_{e}\sin i}^{2}+\sigma_{\log gf}^{2}+\sigma_{cont}^{2}.
\end{equation} 
The mean abundance $<[\frac{X}{H}]>$ is then computed as a weighted mean of the individual abundances [X/H]$_{i}$ derived for each transition (i):\\
\begin{equation}
<[\frac{X}{H}]>=\frac{\sum_{i}([\frac{X}{H}]_{i}/\sigma_{tot_{i}}^{2})}{\sum_{i}(1/\sigma^{2}_{tot_{i}})}
\end{equation}
and the total error, $\sigma$ is given by: \\
\begin{equation}
\frac{1}{\sigma^{2}}=\sum_{i=1}^{N}(1/\sigma_{tot_{i}}^{2})  
\end{equation}
where N is the number of lines per element.\\

\section{Identification of the lines for HD~99803A (HR~4423)}

The lines in the HARPS spectrum of HD~99803A (HR~4423) absorbing more than 2\% of the continuum have been identified using the synthetic spectrum which best fits the observed spectrum.  They are collected in the Table \ref{tab:identifications}.

\startlongtable
\begin{deluxetable}{llllll}
\tabletypesize{\normalsize}  
\tablecaption{Line identifications for HD~99803A (HR~4423)}
\tablecolumns{6}\tablewidth{0pc}  
\tablehead{$\lambda_{\mathrm{obs}}$ (\AA)  & $\lambda_{\mathrm{lab}}$ (\AA) & Identification & $\log gf$ & $E_{low}$ & Relative flux \\}
\startdata
3901.22     & 3900.515    & \ion{Fe}{1}    &  -0.92   & 26140.176 & 0.41  \\
            & 3900.550    & \ion{Ti}{2}    &  -0.45   & 9118.260  &  \\
3903.61     & 3902.945   & \ion{Fe}{1} & -0.47 & 12560.933 & 0.87 \\
3904.42     & 3903.73    & \ion{Fe}{2} & -1.49  &  60807.228  &  0.90 \\
3906.31     & 3905.644  & \ion{Cr}{2}   &  -0.90  & 42986.619  & 0.56 \\
3906.71    & 3906.035  & \ion{Fe}{2}   &  -1.83  & 44929.549  &  0.74  \\
3911.99    & 3911.30   & \ion{Cr}{2}   & -2.06  &  39824.381  &  0.82 \\
3914.14    & 3913.45   & \ion{Ti}{2}   & -0.53  &  8997.710  &  0.43   \\
\enddata
\label{tab:identifications}
\end{deluxetable}

\end{appendix}

\allauthors

\listofchanges

\end{document}